\documentclass[aps,prd,final,reprint,showkeys,showpacs,superscriptaddress,floatfix,nofootinbib,twocolumn,amsmath,amssymb]{revtex4-1}
\usepackage[utf8]{inputenc}

\usepackage{hyperref} 
\usepackage{url}
\usepackage{graphicx}

\usepackage{graphics}
\usepackage{natbib}
\graphicspath{{./}}

\usepackage[normalem]{ulem}
\usepackage{color}

\usepackage{soul}

\usepackage{supertabular}
\usepackage{tabularx}

\def\units#1{~\hbox{$\,{\rm #1}$}}
\def\degrees{\hbox{$^\circ$}}

\begin{document}

\title{Search for dark matter signatures in the gamma-ray emission towards the Sun with the Fermi Large Area Telescope }

\author{M.~N.~Mazziotta}
\email{mazziotta@ba.infn.it}
\homepage{https://orcid.org/0000-0001-9325-4672}
\affiliation{Istituto Nazionale di Fisica Nucleare, Sezione di Bari, via Orabona 4, I-70126 Bari, Italy}
\author{F.~Loparco}
\email{francesco.loparco@ba.infn.it }
\homepage{https://orcid.org/0000-0002-1173-5673}
\affiliation{Istituto Nazionale di Fisica Nucleare, Sezione di Bari, via Orabona 4, I-70126 Bari, Italy}
\affiliation{Dipartimento di Fisica ``M. Merlin" dell'Universit\`a e del Politecnico di Bari, via Amendola 173, I-70126 Bari, Italy}
\author{D.~Serini}
\email{davide.serini@ba.infn.it}
\homepage{https://orcid.org/0000-0002-9754-6530}
\affiliation{Istituto Nazionale di Fisica Nucleare, Sezione di Bari, via Orabona 4, I-70126 Bari, Italy}
\affiliation{Dipartimento di Fisica ``M. Merlin" dell'Universit\`a e del Politecnico di Bari, via Amendola 173, I-70126 Bari, Italy}
\author{A.~Cuoco}
\affiliation{Istituto Nazionale di Fisica Nucleare, Sezione di Torino, via Pietro Giuria 1,  I-10125 Torino, Italy}
\affiliation{Dipartimento di Fisica, Universit\`a di Torino, via P. Giuria 1, 10125 Torino, Italy}
\author{P.~De~La~Torre~Luque}
\affiliation{Istituto Nazionale di Fisica Nucleare, Sezione di Bari, via Orabona 4, I-70126 Bari, Italy}
\affiliation{Dipartimento di Fisica ``M. Merlin" dell'Universit\`a e del Politecnico di Bari, via Amendola 173, I-70126 Bari, Italy}
\author{F.~Gargano}
\homepage{https://orcid.org/0000-0002-5055-6395}
\affiliation{Istituto Nazionale di Fisica Nucleare, Sezione di Bari, via Orabona 4, I-70126 Bari, Italy}
\author{M.~Gustafsson}
\affiliation{Georg-August University G\"ottingen, Institute for theoretical Physics - Faculty of Physics, Friedrich-Hund-Platz 1, D-37077 G\"ottingen, Germany}

\date{\today}

\begin{abstract}
Dark matter particles from the Galactic halo can be gravitationally trapped in the solar core or in external orbits. The enhanced density of dark matter particles either in the solar core or in external orbits can result in the annihilation of these particles producing gamma rays via long-lived intermediate states or directly outside the Sun, respectively. These processes would yield characteristic features in the energy spectrum of the subsequent gamma rays, i.e.,\ a box-like or line-like shaped feature, respectively. We have performed a dedicated analysis using a 10-years sample of gamma-ray events from the Sun collected by the Fermi Large Area Telescope searching for  spectral features in the energy spectrum as a signature of dark matter annihilation. In the scenario of gamma-ray production via long-lived mediators we have also evaluated the dark matter-nucleon spin-dependent and spin-independent scattering cross section constraints from the flux limits in a dark matter mass range from 3\units{GeV/c^2} up to about 1.8\units{TeV/c^2}. In the mass range up to about 150\units{GeV/c^2} the limits are in the range $10^{-46} - 10^{-45}\units{cm^{2}}$ for the spin-dependent scattering and in the range $10^{-48} - 10^{-47}\units{cm^{2}}$ for the spin-independent case. The range of variation depends on the decay length of the mediator.

\end{abstract}


\maketitle

\section{Introduction}
\label{sec:intro}

The existence of dark matter (DM) is well established by several experimental observations~\cite{PhysRevD.98.030001}, while the debate on its nature is still open. Indirect detection of DM annihilation and decay using different astrophysical probes would provide  evidence of the particle nature of DM. Many DM candidate particles are, in fact, expected to annihilate or decay into Standard Model (SM) particles, which can travel from their production sites to the Earth and be detected. All these particles should yield characteristic signals, whose strengths depend on the properties of the parent DM particles (i.e. mass and velocity), on the production process (i.e. DM annihilation/decay cross sections and branching ratios) and on the propagation from the production sites to the Earth. In addition, various astrophysical sources can mimic the signal expected from DM, thus making indirect DM detection highly challenging.

Gamma rays produced in DM annihilations or decays are among the most commonly used probes for indirect DM searches. Unlike charged particles, they are not deflected by the interstellar magnetic fields, thus making directional studies possible, and, unlike neutrinos, they are relatively easy to detect. These properties make gamma rays one of the best probes for DM searches, since they point back to their sources.

The Sun is one of the promising targets for indirect DM searches with gamma rays. DM particles coming from the Galactic halo can interact with the solar nuclei via elastic scattering, being slowed down in each interaction and then captured in the Sun core by its gravitational field. If DM particles can also self-annihilate inside the Sun, an equilibrium between the capture and the annihilation processes will be reached. In the annihilations, different kinds of particles will be produced, but only neutrinos will be able to escape from the Sun. Other particles, such as gamma rays or electron-positron pairs, will be likely absorbed in the Sun and will not be able to reach the Earth. However, some authors~\cite{Pospelov:2007mp,ArkaniHamed:2008qn,Schuster:2009fc,Leane:2017vag,Schuster:2009au,Bell:2011sn,Arina:2017sng} have proposed a possible scenario in which DM particles $\chi$ annihilate into long-lived mediators $\phi$, which are able to escape from the Sun. They will then decay into SM particles, such as gamma rays or electron-positron pairs, that can be detected at Earth.

In another possible scenario, DM particles can scatter inelastically off solar nuclei (see~\cite{TuckerSmith:2001hy,Finkbeiner:2009ug,Chang:2008gd,Menon:2009qj,Nussinov:2009ft,Catena:2018vzc,Blennow:2018xwu}) and be captured after a few interactions. In this model, the captured particles are not able to thermalize and settle into the Sun core, but are orbiting around the Sun. The density of the captured DM particles in external orbits is not negligible, and gamma rays eventually produced in direct DM annihilation can be detected at Earth. 

The Sun is visible in gamma rays because of the interaction of cosmic rays (CRs) with the solar environment. The standard gamma-ray emission of the Sun includes two main components~\cite{Abdo:2011xn,Tang:2018wqp,Seckel:1991ffa,Orlando:2008uk,Orlando:2006zs,mazziotta2020cosmic}: i) the contribution from the disk, originating from the interactions of hadronic CRs with the solar atmosphere, which yield hadronic cascades with their gamma-ray component; ii) the contribution from diffuse emission, due to the interactions of cosmic-ray electrons and positrons with the optical solar photons in the heliosphere, in which gamma rays are produced by inverse Compton scattering.

The standard solar emission mechanisms are expected to yield a smooth gamma-ray spectrum, while both the DM scenarios illustrated above are expected to yield some characteristic features in the spectrum.
In particular, if gamma rays are produced through a mediator, the spectrum should exhibit a box-like feature~\cite{Mazziotta:2019CREs,Ibarra:2012dw}; on the other hand, if gamma rays are produced directly in DM annihilations, a line-like feature is expected~\cite{Sivertsson:2009nx}. 

The Large Area Telescope (LAT) onboard the Fermi Gamma-Ray Space Telescope routinely observes the Sun during its data taking. In the present work we have implemented a dedicated analysis based on a Poisson maximum likelihood approach to search for a possible DM signature in the energy spectrum of solar gamma rays. We use a 10-year dataset of gamma ray events collected by the Fermi-LAT and analyzed with the newest Pass 8 event selection in an observed energy range from $100\units{MeV}$ up to $150 \units{GeV}$. 

We have implemented a dedicated analysis technique, combining the data from two Regions of Interest (RoIs), centered respectively on the Sun and on a “time-offset” Sun position (hereafter anti-Sun). The anti-Sun corresponds to a sky position which is always well separated from the Sun: in fact, since we choose a 6 months forward/backward time offset, the angular separation between the Sun and the anti-Sun is close to $180\degrees$. In this approach, the anti-Sun RoI is used as a control region to take into account any possible systematic uncertainties.

\section{Solar Dark Matter models}
\label{sec:models}

In this work we investigate the two possible scenarios outlined in the Introduction, in which gamma rays are produced in the annihilations of DM particles captured in the Sun. In both scenarios, the equilibrium between DM capture and annihilation is regulated by the following equation:

\begin{equation}
\label{eq:balance}
\frac{dN_\chi }{dt} = \Gamma_{\text{cap}} – C_{\text{ann}} N_{\chi}^2 
\end{equation}
where $N_\chi(t)$ is the number of DM particles in the Sun at a given time $t$, $\Gamma_{\text{cap}}$ is the DM capture rate and $C_{\text{ann}}$ is the DM annihilation factor, which depends on the annihilation cross section.   
In this equation we have not considered the evaporation mechanism, which is not relevant for DM masses above a few$\units{GeV}$~\cite{Albert:2018jwh,griest1987cosmic}.

When equilibrium is reached ($dN_\chi/dt = 0$), the annihilation rate is independent of the annihilation cross section and is set by the capture rate $\Gamma_{\text{cap}}$, which in turn depends on the scattering cross section (either spin-dependent, $\sigma_{\text{SD}}$, or spin independent, $\sigma_{\text{SI}}$), on the local halo DM number density $\rho_{\odot}$, on the DM mass $m_\chi$, on the DM velocity distribution and on its dispersion:

\begin{equation}
\label{eq:caprate}
 \Gamma_{\text{ann}} = \frac{1}{2} C_{\text{ann}}N_\chi^2 = \frac{1}{2} \Gamma_{\text{cap}}
\end{equation}
The factor $1/2$ accounts for the two DM particles involved in each annihilation event.

In this section we briefly review both processes of DM particle capture and annihilation in the Sun and the evaluation of the fluxes of the gamma rays at Earth in both cases. More detailed description of the models used in this work can be found in Ref.~\cite{Mazziotta:2019CREs}.

If equilibrium between capture and annihilation is not reached, the solution of eq.~\ref{eq:balance} is given by:

\begin{equation}
N_{\chi}(t) = \sqrt{\frac{\Gamma_{cap}}{C_{ann}}} \tanh \left( \frac{t}{\tau} \right) \end{equation}
where $\tau=(\Gamma_{cap}C_{ann})^{-1/2}$~\cite{Jungman:1995df} is the equilibrium time scale of the process. As a consequence, the annihilation rate in eq.~\ref{eq:caprate} should be multiplied by the factor $\tanh^{2}(t/\tau)$. In Section~\ref{sec:res} we will evaluate the limits on the DM-nucleon cross section under the assumption of equilibrium between capture and annihilation, while in Section~\ref{sec:con} we will discuss how these limits change in the non-equilibrium case. 

\subsection{Annihilation through a light intermediate state}
\label{sec:inter}

In the first scenario, DM particles are captured by the Sun through elastic scattering interactions with the solar nuclei. DM particles lose their energy in subsequent interactions until they sink into the core of the Sun, reaching thermal equilibrium. In the framework of this model, we assume that DM particles annihilate into a light intermediate state $\phi$, which subsequently decays into photons through the processes $\chi \chi \rightarrow \phi \phi$ and $\phi \rightarrow \gamma \gamma$. In this picture we implicitly assume that the mediators $\phi$ are able to escape the Sun without further interactions and that, after escaping from the Sun, each $\phi$ decays into a pair of gamma rays. If this happens, the gamma rays produced can reach the Earth and may be detectable as a signature of DM in the form of an excess of photons from the direction of the Sun.

The DM particles are assumed to annihilate at rest in the Sun core, and in the lab frame the energy of the mediator $\phi$ will be equal to the mass $m_\chi$ of the DM particle, i.e. $E_\phi = m_\chi$.
If we also assume that the $\phi$ is a light mediator such that $m_\phi \ll m_\chi$, the angular dispersion of the daughter gamma rays with respect to the direction of the parent $\phi$ will be negligible. Under these assumptions, the gamma-ray flux will be equivalent to that from a point-like source centered in the Sun’s core.

The DM gamma-ray flux at Earth is given by~\cite{Mazziotta:2019CREs}:

\begin{equation}
\Phi_{\text{DM}}(E) =  N_{\gamma}(E) \frac{\Gamma_{\text{cap}}}{4 \pi D^2} \left( e^{-R_\odot/L} - e^{-D/L} \right) 
\label{eq:phidm}
\end{equation}
where $R_{\odot}$ is the solar radius, $D$ is the Sun-Earth distance and $L$ is the $\phi$ decay length. The gamma-ray flux at Earth is therefore equivalent to the flux from a point–like source in the Sun ($\Gamma_{\text{cap}}/4 \pi D^2$) modulated by the survival probability of the mediator $\left( e^{-R_\odot/L} - e^{-D/L} \right)$. Under the assumptions of this model, the resulting DM photon spectrum will have a box-shape, where the center and the width of the box will depend on $m_{\chi}$ and $m_{\phi}$~\cite{Ibarra:2012dw}. In the hypothesis of a light mediator, i.e. $m_{\phi}\ll$ $m_{\chi}$, the dependence on $m_{\phi}$ is lost and the box extends from $E=0$ to $E=m_\chi$. In this case $N_{\gamma}(E)$ can be written as $N_{\gamma}(E)=2 H (m_\chi-E) /m_\chi$, where $H$ is the Heaviside step function and the factor $2$ takes into account the fact that for each mediator decay two photons are produced.

\begin{figure}[!ht]
\includegraphics[width=\columnwidth,height=0.27\textheight,clip]{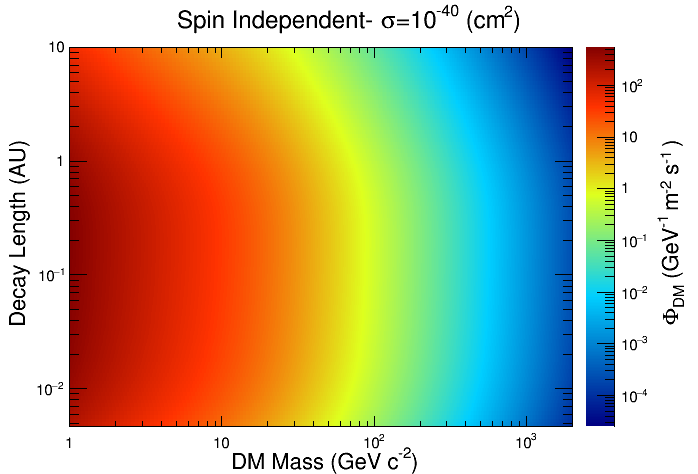}
\includegraphics[width=\columnwidth,height=0.27\textheight,clip]{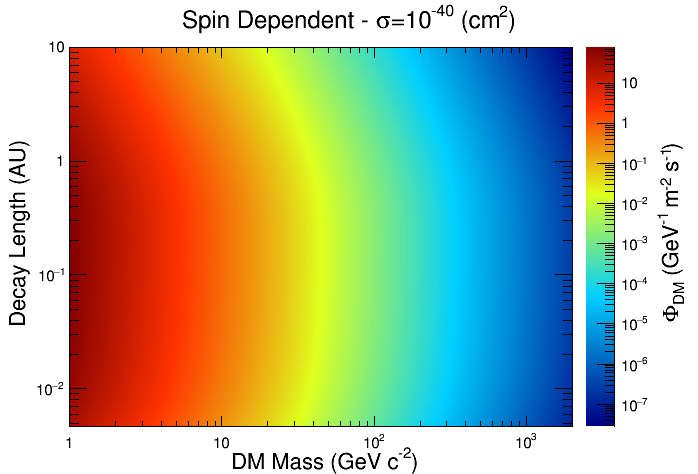}
\caption{ Expected gamma-ray flux at the Earth from DM annihilations in the Sun via long-lived mediators as a function of the DM mass and decay length, for the spin independent (top panel) and spin dependent (bottom panel) cases. In both cases a scattering cross section of $10^{-40}\units{cm^{2}}$ is assumed.}
\label{fig:dmflux}
\end{figure}

In this work the capture rate $\Gamma_{\text{cap}}$ has been calculated with the {\tt DARKSUSY} code version 6.1.0~\cite{Gondolo:2004sc,Bringmann:2018lay,darksusyweb} assuming the default settings, i.e. a local DM density $\rho_\odot = 0.3\units{ GeV/cm^3}$, a DM-nucleon scattering cross section $\sigma=10^{-40}\units{cm^2}$ (in both the spin-dependent and spin-independent cases) and a Maxwellian velocity distribution for DM particles with average $v_\odot=220\units{km/s}$ and dispersion $v_{rms}=270\units{km/s}$. A change in these values will result in a rescaling of the total capture rate. In particular, the dependence of the capture rate on the DM-nucleon cross section is linear. The capture rates for both the spin-dependent and spin-independent cases are shown in Fig.2 of Ref.~\cite{Mazziotta:2019CREs}.
Fig.\ref{fig:dmflux} shows the expected gamma-ray fluxes at Earth from DM annihilations in the Sun core via long-lived mediators as a function of the DM mass and decay length.

\subsection{Inelastic dark matter}
\label{sec:idm}

In the previous scenario DM particles lose energy through subsequent elastic scatterings, sinking to the core of the Sun. In this case, the fraction of DM particles captured outside the solar surface is negligible~\cite{Sivertsson:2009nx}. In the second scenario considered in this work, DM particles undergo inelastic scatterings with solar nuclei in which an excited state $\chi^{*}$ is produced with a slightly heavier mass than the $\chi$, i.e. $\chi + N \rightarrow \chi^{*} + N$.
In this model, each DM particle can be scattered only if its energy is larger than a threshold  $E_{thr}=\Delta (1+m_\chi/m_N)$, where $\Delta = m_\chi^{*} - m_\chi$ is the mass splitting parameter and $m_N$ is the mass of the nucleus. Hence, DM particles captured by the Sun can undergo only a few interactions, in which they are slowed down until their energy reaches the threshold; when this happens, DM particles cannot thermalize in the Sun core. 
If the elastic scattering cross section is sufficiently small ($\sigma < 10^{-47} \units{cm^2}$), the elastic capture mechanism will be inefficient and particles will not be able to sink to the Sun's core. Thus, the density of DM particles outside the Sun in external orbits will remain significant.
As a consequence, the annihilations of these trapped DM particles could yield an observable flux of gamma rays from the Sun direction.

Since in this scenario the DM density is expected to decrease rapidly with increasing distances from the Sun, we consider only annihilations occurring close to the solar surface. We also assume that the DM particles annihilate at rest, so that the energy of their daughter gamma rays is given by $E_{\gamma}=m_{\chi}$. 
Under these assumptions, we expect that the energy spectrum of solar gamma rays detected at Earth should exhibit a line-like feature, which would represent the DM signature for this scenario.

\section{Data selection}
\label{sec:data}
The analysis presented in this work has been performed using the {\tt Pass 8} data~\cite{Atwood:2013rka} (specifically P8R3\_CLEAN photon events~\footnote{We decided to use P8R3\_CLEAN photon events because this selection is recommended for the analysis of sources outside the Galactic plane~\cite{Bruel:2018lac,cicerone} and, as it will be discussed later, in our analysis we select time intervals when the Sun is away of at least $5\degrees$ from the Galactic plane.}) collected by the Fermi LAT during its first 10 years of operation. We search for possible features in the gamma-ray spectrum towards the Sun, which should appear as photon excesses with respect to the steady solar emission in the region above $100\units{MeV}$. 

We have implemented an analysis based on an on/off technique, in which we combine the data from a Region of Interest (RoI) centered on the position of the Sun (hereafter ``on region'') and from a RoI centered on the anti-Sun (hereafter ``off region''). A crucial point in any analysis of gamma rays from the Sun is the treatment of the background. 
The background sources include the diffuse gamma-ray emission, the point sources eventually encountered by the Sun along its path in the sky, those charged cosmic rays being misclassified as photons and all the irreducible background (e.g., gamma rays detected by the LAT that originate in cosmic-ray interactions with the spacecraft). Since the Sun is a moving source, the background changes with time, and building an appropriate template background model would be not easy. On the other hand, the use of a control region (the ``off'' region) allows direct evaluation of the background from the data. In addition, this approach allows direct accounting for any possible systematic effects that arise in both the on and off regions. 

Given that, (i) the Sun is seen from Earth as a disk with an angular radius of about $0.25\degrees$, (ii) photons from solar DM should yield a signal peaked towards the direction of the Sun~\cite{Sivertsson:2009nx}, and (iii) photons produced from the decay of a light mediator traveling from the Sun to the Earth would deviate by a few degrees at most from the direction of the Sun, we have defined the ``on'' and the ``off'' regions as disks of $2 \degrees$ angular radius centered on the Sun and on the anti-Sun respectively. 

At any given time, the position of the anti-Sun is defined as the position that the Sun will take 6 months later; if the offset time exceeds the maximum time of the data set used in the analysis, the time-offset position will be brought back to the position taken by the Sun at the beginning of the time interval analyzed; the anti-Sun will then follow the path of the Sun during the first $6$ months of the interval.  In this way the ``off'' region will span the same portion of sky as the ``on'' region, and the anti-Sun will be always separated by about $180\degrees$ from the Sun. The position of the Sun is obtained from its ephemeris using a software interface to the JPL libraries\footnote{\url{https://ssd.jpl.nasa.gov/horizons.cgi}}. 
For the analysis of the ``on'' (and ``off'') regions, we restricted the time intervals (``good time intervals", GTIs) to the period when the LAT was operating in its standard science operation configuration and was outside the South Atlantic Anomaly (SAA). We have also excluded from the analysis the time intervals in which solar flares were registered. To avoid contamination from the bright limb of the Earth we also discarded the data taken during the times in which the separation between the center of the ``on'' (``off'') RoI direction and the zenith exceed $78\degrees$. We also selected GTIs when the RoIs were observed with off-axis angles in the instrument frame smaller than $64.5\degrees$. To mitigate systematic uncertainties due to the bright diffuse gamma-ray emission from the Galactic plane, we selected only the times when the Sun (anti-Sun) was at least $5\degrees$ away from the Galactic plane, i.e., times when the latitude of the Sun is \mbox{$|b|>5\degrees$}. Finally, we also required a minimum angular separation of $4\degrees$ between the Sun (anti-Sun) and the Moon or  any bright\footnote{Here we define as ``bright'' a source whose gamma-ray flux above $100 \units{MeV}$ is larger than $4\times 10^{-7} \units{photons~cm^{-2}~s^{-1}}$.} celestial source in the 3FGL Fermi LAT source catalog~\citep{Acero:2015hja}. 

\section{Analysis procedure}
\label{sec:ana}
In our analysis we have implemented a Poisson maximum likelihood fitting procedure in order to search for possible local excesses in the count spectra of photons from the ``on'' region. The fits are performed in sliding energy windows to search for DM signatures, which should yield local excess counts on the top of a smooth spectrum. The excesses in the count spectrum are assumed to originate either from a delta-like or a box-like feature in the energy spectrum~\cite{Mazziotta:2017ruy,Mazziotta:2019CREs}. When folded with the instrument energy response function, these features should appear as a broad peak and a smooth edge respectively, with a width determined by the energy resolution of the LAT. 

Together with the count spectrum of the ``on'' region, we also fit the spectrum from the ``off'' region. With this approach, possible systematic effects are  automatically included in the analysis, since they should appear in both the ``on'' and ``off'' regions. 

The observed energy range scanned extends from $100\units{MeV}$ to $150\units{GeV}$. Each energy window is defined as the interval $[(1-w)E_{\text{w}},(1+w)E_{\text{w}}]$, where $E_{\text{w}}$ is the energy corresponding to the center of the window and $wE_{\text{w}}$ is the half-width of the window. The parameter $w$ is chosen in order to ensure that the width of the windows is larger than the LAT energy resolution for the whole energy range explored in the analysis. Different window sizes have been tested as will be discussed in section~\ref{sec:res}. The total energy range is divided in 64 bins per decade, equally spaced on a logarithmic scale. 

We assume that the observed counts in each energy bin of the ``on'' and ``off'' regions are independent and Poisson distributed. Indicating with $\vec{n}^{on}=(n_{1}^{on}, n_{2}^{on}, \ldots n_{\text{N}}^{on})$ and $\vec{n}^{off}=(n_{1}^{off}, n_{2}^{off}, \ldots n_{\text{N}}^{off})$ the counts detected in the $N$ energy bins in the ``on'' and in the ``off'' regions respectively, and with $\vec{\mu}^{on}=(\mu_{1}^{on}, \mu_{2}^{on}, \ldots \mu_{N}^{on})$ and  $\vec{\mu}^{off}=(\mu_{1}^{off}, \mu_{2}^{off}, \ldots \mu_{N}^{off})$ the corresponding expected counts, we define the likelihood functions for the two regions as:

\begin{eqnarray}
\mathcal{L}_{on}(\vec{n}^{on} | \vec{\mu}^{on}) & = & \prod_{j=1}^{N} \rm{e}^{-\mu_{\text{j}}^{on}} \frac{(\mu_{\text{j}}^{on})^{n_{\text{j}}^{on}}}{n_{\text{j}}^{on}!} \\
\mathcal{L}_{off}(\vec{n}^{off} | \vec{\mu}^{off}) & = & \prod_{j=1}^{N} \rm{e}^{-\mu_{\text{j}}^{off}} \frac{(\mu_{\text{j}}^{off})^{n_{\text{j}}^{off}}}{n_{\text{j}}^{off}!}.
\end{eqnarray}
As will be discussed in the following part of this section, we have implemented several fits, in which we have maximized the global likelihood, defined as $\mathcal{L}=\mathcal{L}_{on} \times \mathcal{L}_{off}$.

The expected counts in the \textit{$i$-th} bin of reconstructed energy of the two regions are given by:

\begin{equation}
\mu_{\text{i}}^{on}  =  t_{on} \int dE_{\text{t}} \mathcal{R}_{on}(E_{i}|E_{\text{t}}) \left[ \Phi_{\text{sig}}(E_{\text{t}})+\Phi_{\text{bkg}}(E_{\text{t}}) \right]
\label{eq:mus} 
\end{equation}
\\
\begin{equation}
\mu_{\text{i}}^{off} =  t_{off} \int dE_{\text{t}} \mathcal{R}_{off}(E_{\text{i}}|E_{\text{t}}) \Phi_{\text{bkg}}(E_{\text{t}}) 
\label{eq:mub} 
\end{equation}

In the previous equations we indicate with $E_{i}$ the reconstructed energy corresponding to the \textit{$i$-th} bin and with $E_{\text{t}}$ the true photon energy; $t_{on/off}$ are the integrated livetimes of the two regions; $\mathcal{R}_{\text{on/off}}(E_{\text{i}}|E_{\text{t}})$ are the instrument response matrices incorporating the effective area, the angular resolution~\footnote{When evaluating the instrument response matrices from the Monte Carlo simulation we select photon events with an angular separation between the reconstructed and the true photon directions less than the angular radius of the RoI, i.e. $2\degrees$.} and the energy resolution of the LAT~\cite{Mazziotta:2009rd,Loparco:2009by}, evaluated by means of the Monte Carlo (MC) simulations~\cite{Ackermann:2012kna}, taking into account the livetime distributions as a function of the off-axis angle in the instrument frame; finally, $\Phi_{\text{sig}}(E_{\text{t}})$ is the flux model for the Sun, which is taken to contribute only in the ``on'' region, while $\Phi_{\text{bkg}}(E_{\text{t}})$ is the flux model for the background (diffuse, point sources, irreducible background), which is assumed to contribute equally in the ``on'' and ``off'' regions.

Each flux model contribution can be expressed as the sum of a continuous smooth component $\Phi_{0}(E)$ with a possible additional feature $\Phi_{\text{f}}(E)$. Since the energy windows are narrow, the continuous term $\Phi_{0}(E)$ can be well described as a simple power law model (PL) $\Phi_{0}(E; k, \alpha) = k (E/E_{0})^{-\alpha}$,  where $\alpha$ is the spectral index and the prefactor $k$ (in units of \units{MeV^{-1} cm^{-2} s^{-1}}) corresponds to the photon flux at the scale energy $E_{0}$, fixed at $100 \units{MeV}$. We assume two different models for the spectral feature $\Phi_{\text{f}}(E; s)$: a delta-like (line) feature, \mbox{$\Phi_{\text{f}}(E; s) = s ~ \delta(E_{\text{f}}-E)$}, and a box-like (box) feature \mbox{$\Phi_{\text{f}}(E; s) = s ~ H(E_{\text{f}}-E)$}, where $E_\text{f}$ is the characteristic energy of the feature (either the line energy or the upper edge of the box), $\delta$ is the Dirac delta function, $H$ is the Heaviside step function and $s$ represents the intensity of the feature in units of \units{cm^{-2}~s^{-1}} for the line and of \units{GeV^{-1}~cm^{-2}~s^{-1}} for the box. If $E_{\text{f}}(1+w) < 150\units{GeV}$, the search for the feature is performed in the window centered on $E_{\text{f}}$; if $E_{\text{f}}(1+w)>150\units{GeV}$, the fits are performed in the last window, i.e. the window with upper bound at $150\units{GeV}$. 
We include a feature in the background flux model with the same spectral shape as the feature in the signal flux model to account for possible systematic effects which could mimic a real feature in the solar spectrum. A feature arising from an instrumental effect (e.g. incorrect energy reconstruction) should in fact appear in the count spectra of both the ``on'' and ``off'' regions.       

We test the hypothesis of a DM-induced gamma-ray signal against the null hypothesis in which no DM signals are included. In the null hypothesis model ($H_0$) we include the feature only in the background flux model, while we assume a continuous signal flux, i.e. $\Phi_{\text{bkg}}(E)=k_{\text{bkg}} (E/E_{0})^{-\alpha_{\text{bkg}}} + \Phi_{\text{f,bkg}}(E; s_{\text{bkg}})$ and $\Phi_{\text{sig}}(E|H_0)=k_{\text{sig}} (E/E_{0})^{-\alpha_{\text{sig}}}$. In the alternative hypothesis model ($H_1$) we include a feature also in the signal flux model, i.e. we assume $\Phi_{\text{sig}}(E|H_1)=k_{\text{sig}} (E/E_{0})^{-\alpha_{\text{sig}}} + \Phi_{\text{f,sig}}(E; s_{\text{sig}})$ with $s_{sig}>0$.\footnote{We require $s_{sig}>0$ since a possible feature in the signal model appears as an excess with respect to the continuous signal flux. On the other hand, the possible features in the background model can appear either as a flux excess or as a flux deficit, since they originate from instrumental effects.}
The parameters of each fit are evaluated using the {\tt MINUIT} code within the ROOT toolkit~\cite{Brun:1997pa,rootweb}; {\tt MIGRAD} is used as minimization algorithm.

The goal of this analysis is to test whether a signal feature is significantly observed in the  photon spectrum from the Sun. In each energy window we evaluate the local Test Statistic $TS = - 2 (log\mathcal{L}_{H_{0},max} - log\mathcal{L}_{H_{1},max})$, where $\mathcal{L}_{H_{0},max}$ and $\mathcal{L}_{H_{1},max}$ are the maximum values of the likelihood functions obtained when fitting the data with the models corresponding to the null hypothesis and the alternative hypothesis, respectively. Since the two models are nested and the null hypothesis model can be obtained from the alternative hypothesis model setting $s_{\text{sig}}=0$, we expect that $TS$ should obey a $\chi^{2}$ distribution with one degree of freedom. 

To evaluate the upper limit (UL) at $95\%$ confidence level on the parameter $s_{\text{sig}}$, we have searched for the value which gives  $log \mathcal{L}(s_{\text{sig}}) = log\mathcal{L}_{H_{1},max}-2.71/2$ using the processor {\tt MINOS} in {\tt MINUIT}.

Indeed, for both the box and line cases, we have seen that in the energy region below 1\units{GeV} the local $TS$ distributions exhibit small deviations from the expected distribution. Nonetheless we have also seen, using the pseudo-experiment technique discussed below, that the confidence intervals evaluated using the actual distribution of $TS$ do not differ significantly from those evaluated using the expected $\chi^{2}/2$ distribution, and therefore we have decided to quote the limits evaluated with the standard approach. 

To evaluate the expectation bands for our results, i.e. the sensitivity to the null hypothesis, we use a pseudo-experiment technique. As starting point, we fit the observed count distributions assuming that the fluxes in both the ``on'' and ``off'' regions can be modeled in the whole energy range with a smoothly broken power law:

\begin{equation}
\Phi =  k\Bigg(\frac{E}{E_{\text{b}}}\Bigg)^{-\alpha_{1}}\Bigg[1+\Bigg(\frac{E}{E_{\text{b}}}\Bigg)^{\frac{1}{\beta}}\Bigg]^{(\alpha_1 - \alpha_2)\beta}
\label{eq:over} 
\end{equation}
where $E_{b}$ is the energy break, $\alpha_{1,2}$ are the spectral indices below and above the break and the parameter $\beta$ sets the smoothness of the slope change. This model has been chosen in order to best represent the overall trend of the experimental counts. All parameters, i.e. the energy break, the slopes and the smoothness parameter $\beta$, have been fitted. The values of the fitted parameters using $2\degrees$ RoIs for the two regions are summarized in Tab.\ref{tab:over}.

\begin{table}[h]
\resizebox*{0.90\columnwidth}{!}{
\begin{tabular}{lcc}
&\bf{``off'' region}  &\bf{``on'' region} \\ \toprule
{$k$ (\rm{MeV$^{-1}$ cm$^{-2}$ s$^{-1}$})} &$(5.14\pm0.42)\times10^{-11}$ &$(5.58\pm0.03)\times10^{-10}$\\
{$E_{\text{b}}$ (\rm{MeV})}  &$872\pm30$ &$469\pm13$      \\    
{$\alpha_{1}$}    &$1.60\pm0.01$ &$1.05\pm0.02$    \\       
{$\alpha_{2}$}    &$2.79\pm0.03$ &$2.47\pm0.01$   \\     
{$\beta$}        &$0.87\pm0.37$  &$0.80\pm0.05$     \\  
\hline\hline
\end{tabular}
}
\caption{“Best-fit parameters for a broken power-law template for the fluxes in the ``off'' and ``on'' regions (Eq.\ref{eq:over}). The angular radius of both regions is $2\degrees$.}
\label{tab:over}
\end{table}

We then perform a set of $1000$ pseudo-experiments, in which the counts in each energy bin of the ``on'' and ``off'' regions are extracted from Poisson distributions with mean values obtained from the template flux models (Eq.~\ref{eq:over}). Finally, we apply the fit procedure described above to each pseudo-experiment and we evaluate the distributions of the fitted parameters and of the $TS$. The containment bands for the parameters and for the $TS$ are then evaluated as the quantiles of these distributions.

\begin{figure}[!ht]
    \centering
    \includegraphics[width=\columnwidth,height=0.25\textheight,clip]{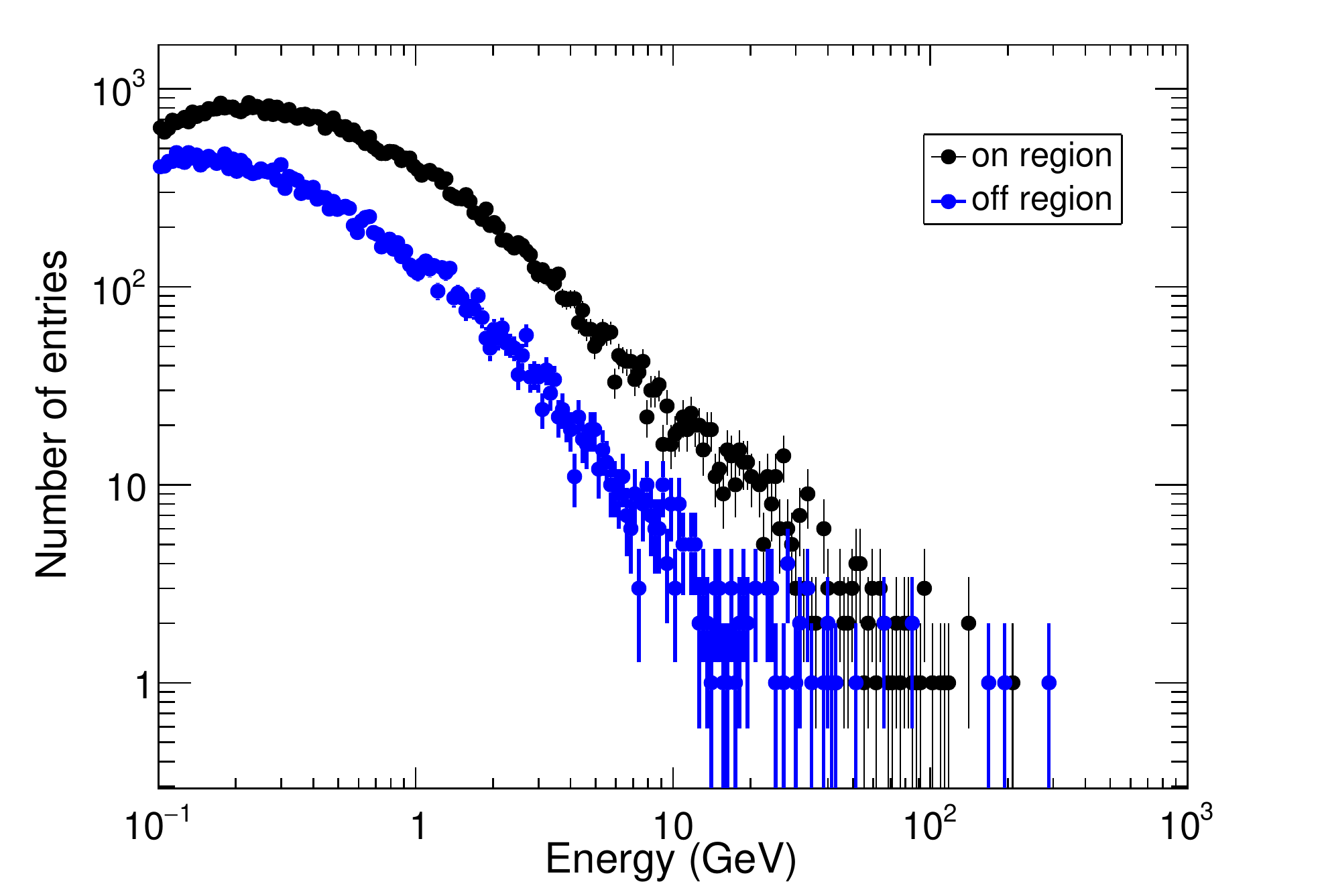}
    \caption{Observed count distributions in the ``on'' and ``off'' regions, both within an angular radius of $2\degrees$ and $64$ energy bin per decade. 
    The error bars correspond to the square roots of the counts in each bin.}
    \label{fig:counts}
\end{figure}

\section{Results}
\label{sec:res}

In Figure~\ref{fig:counts} the observed count distributions in the ``on'' and ``off'' regions are shown respectively in black and blue using the RoI of $2\degrees$ of angular radius. Since the exposures of the two regions are nearly equal, the excess of counts in the ``on'' region is due to the steady solar gamma-ray emission. 

\begin{figure*}[!ht]
    \includegraphics[width=\columnwidth,height=0.22\textheight,clip]{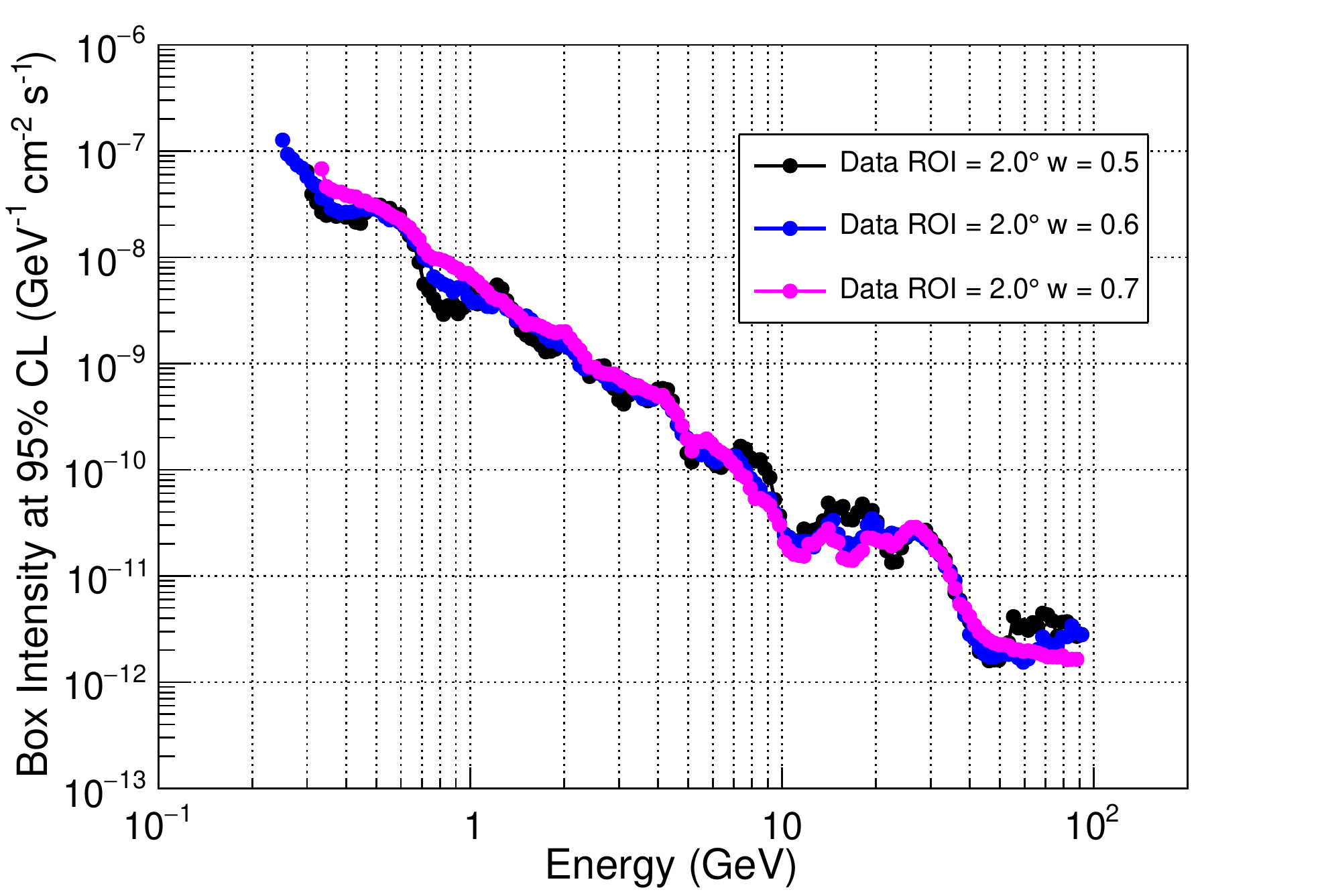}
    \includegraphics[width=\columnwidth,height=0.22\textheight,clip]{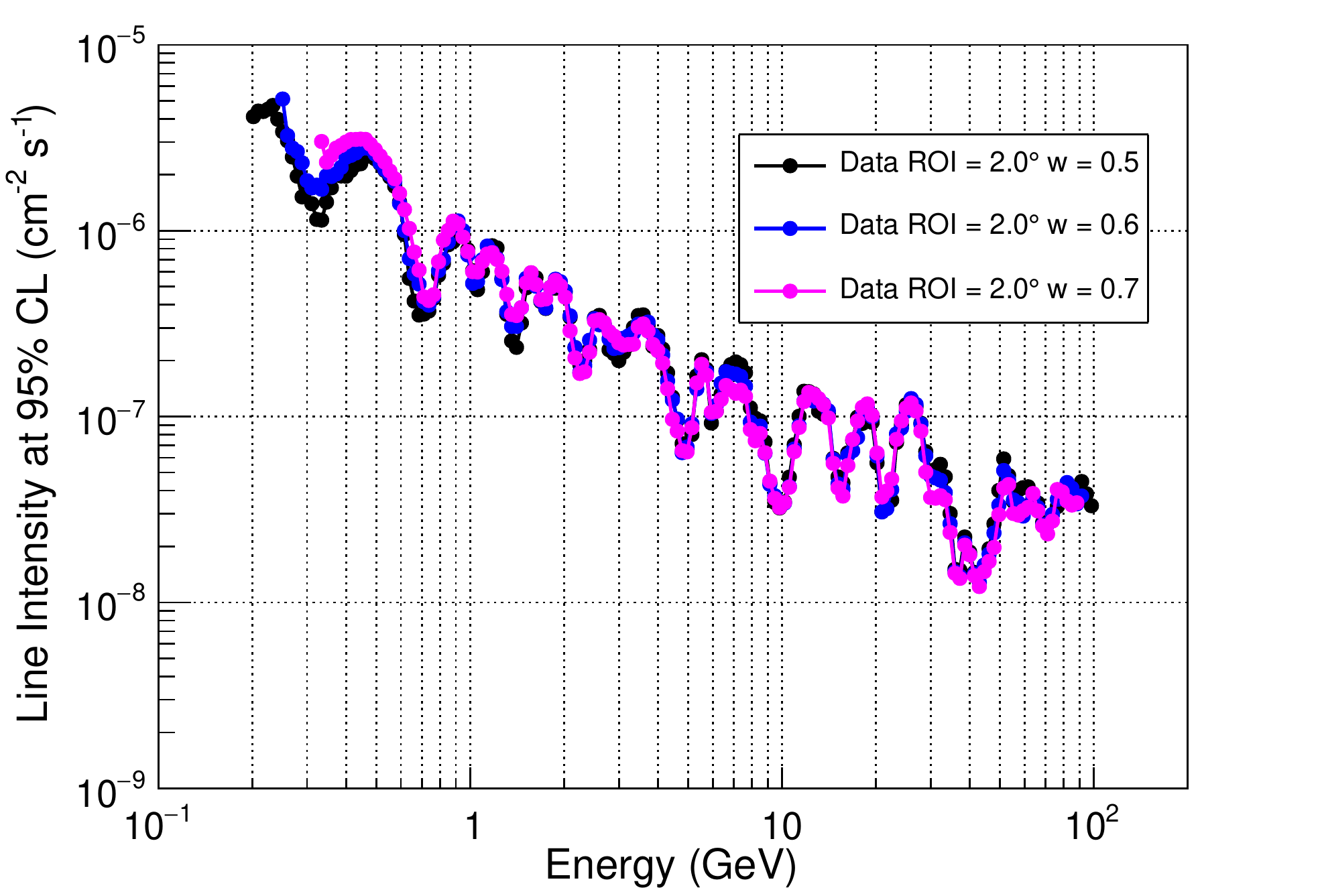}
        \caption{Upper limits for fits performed using a RoI of $2\degrees$ angular radius and different energy window sizes for the box-like features (left plot) and for the line-like features (right plot).}
    \label{fig:wComp}
\end{figure*}

The size $w$ of the energy window to be used in the fit should take into account the energy resolution of the instrument, which spreads the features over several energy bins. The LAT energy resolution for the photon events at $95\%$ containment is about $0.40$ at $100\units{MeV}$ (the minimum energy chosen for our analysis) and decreases with increasing energy (about $0.20$ at $10\units{GeV}$)~\cite{Atwood:2013rka,P8performance}. 

To test the dependence of the results on the choice of the energy window size, we have applied the same analysis procedure increasing the value of $w$ above the worst energy resolution. Figure~\ref{fig:wComp} shows a comparison among the upper limits obtained with different window sizes, i.e. $w=0.5$, $0.6$ and $0.7$, for both the box-like (left plot) and the line-like features (right plot). The values of the upper limits are almost independent of the window size in both cases.
For the present analysis we choose to use a window size of $0.6$.

We have also tested the dependence of the results on the choice of the ``off'' region. We have applied our analysis procedure using the same ``on'' region defined in Section~\ref{sec:data} and different ``off'' regions. The first set of tests has been performed changing the angular radius of the ``off'' region, from the nominal value of $2\degrees$ up to $4\degrees$. We have then performed a second set of tests, in which we have selected ``off'' regions of $2\degrees$ angular radius, with different time offsets with respect to the nominal ``off'' region ($\pm 1$ and $\pm 3$ months). Finally, we have repeated our analysis combining multiple ``off'' regions, corresponding to different time-offset positions of the Sun. For both the searches of box-like and line-like features we see that the results do not change when changing the ``off'' region and we can therefore conclude that the results are robust.

\begin{figure*}[!ht]
\includegraphics[width=\columnwidth,height=0.25\textheight,clip]{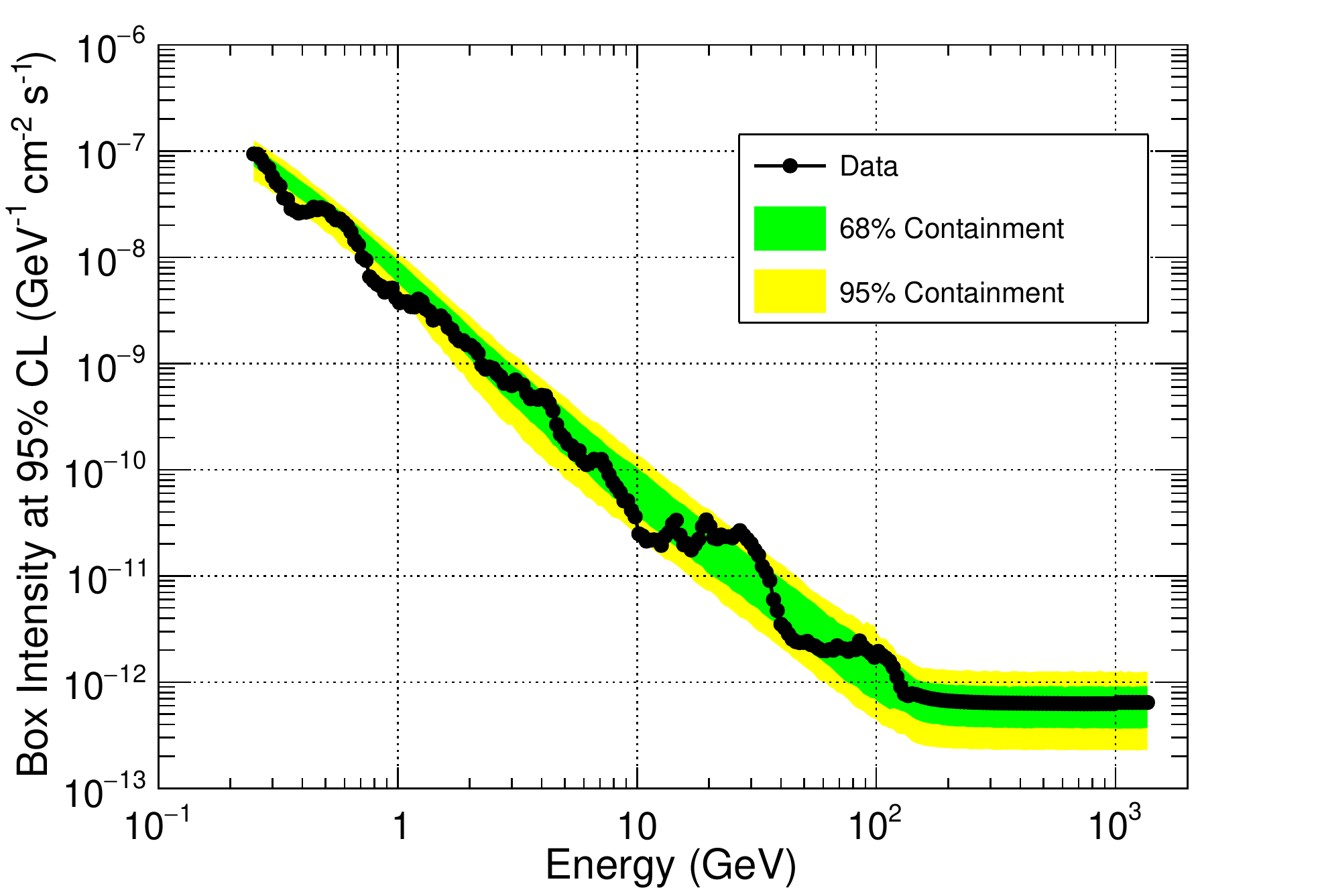}
\includegraphics[width=\columnwidth,height=0.25\textheight,clip]{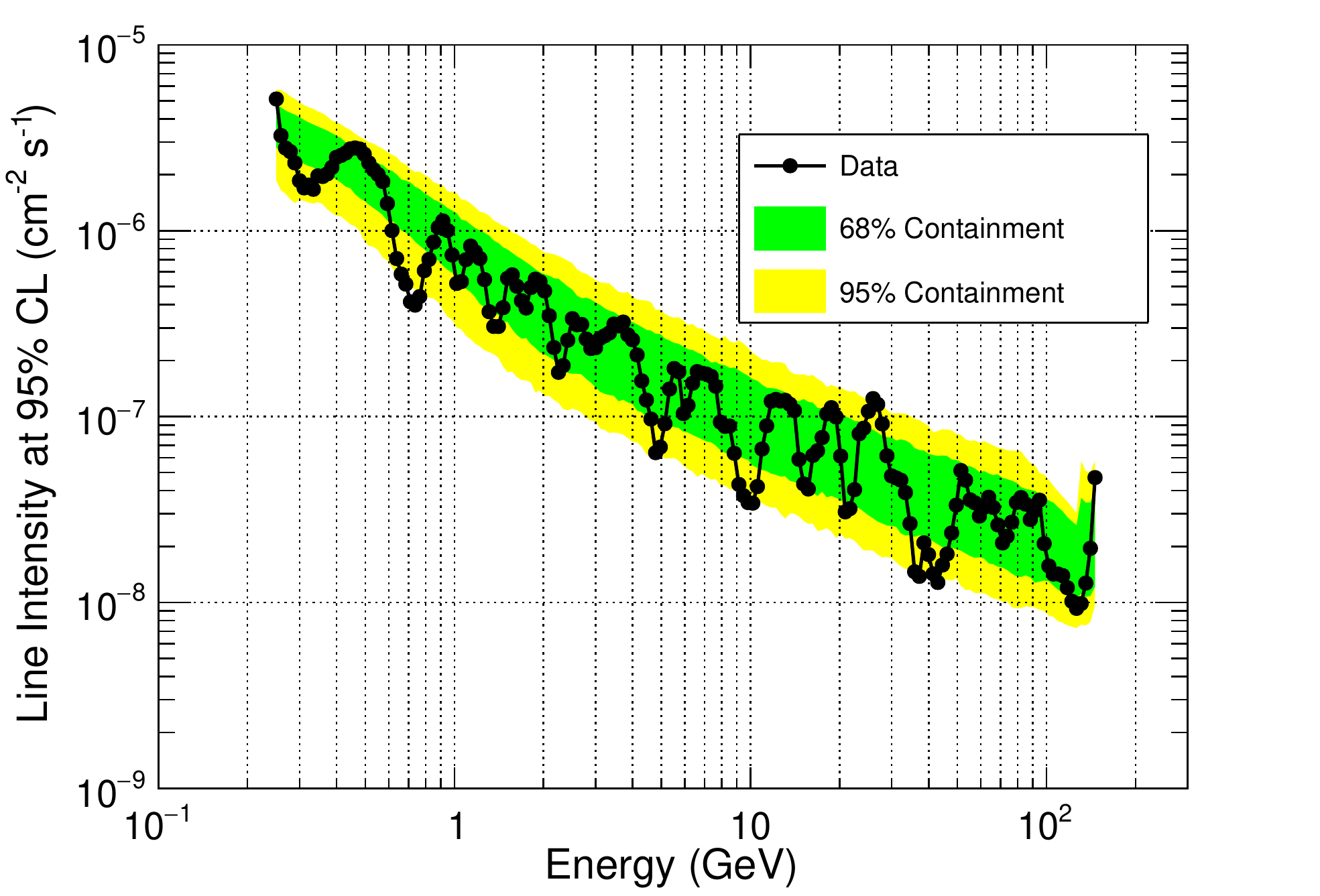}
\caption{Upper limit at 95\% confidence level on the intensity of the feature in the solar spectrum. Left panel: box model; Right panel: line model. The green and yellow regions indicate the central 68\% and 95\% expectation bands for the 95\% CL limits evaluated from the pseudo-experiments.}
\label{fig:ul}
\end{figure*}

\begin{figure*}[!ht]
\includegraphics[width=\columnwidth,height=0.25\textheight,clip]{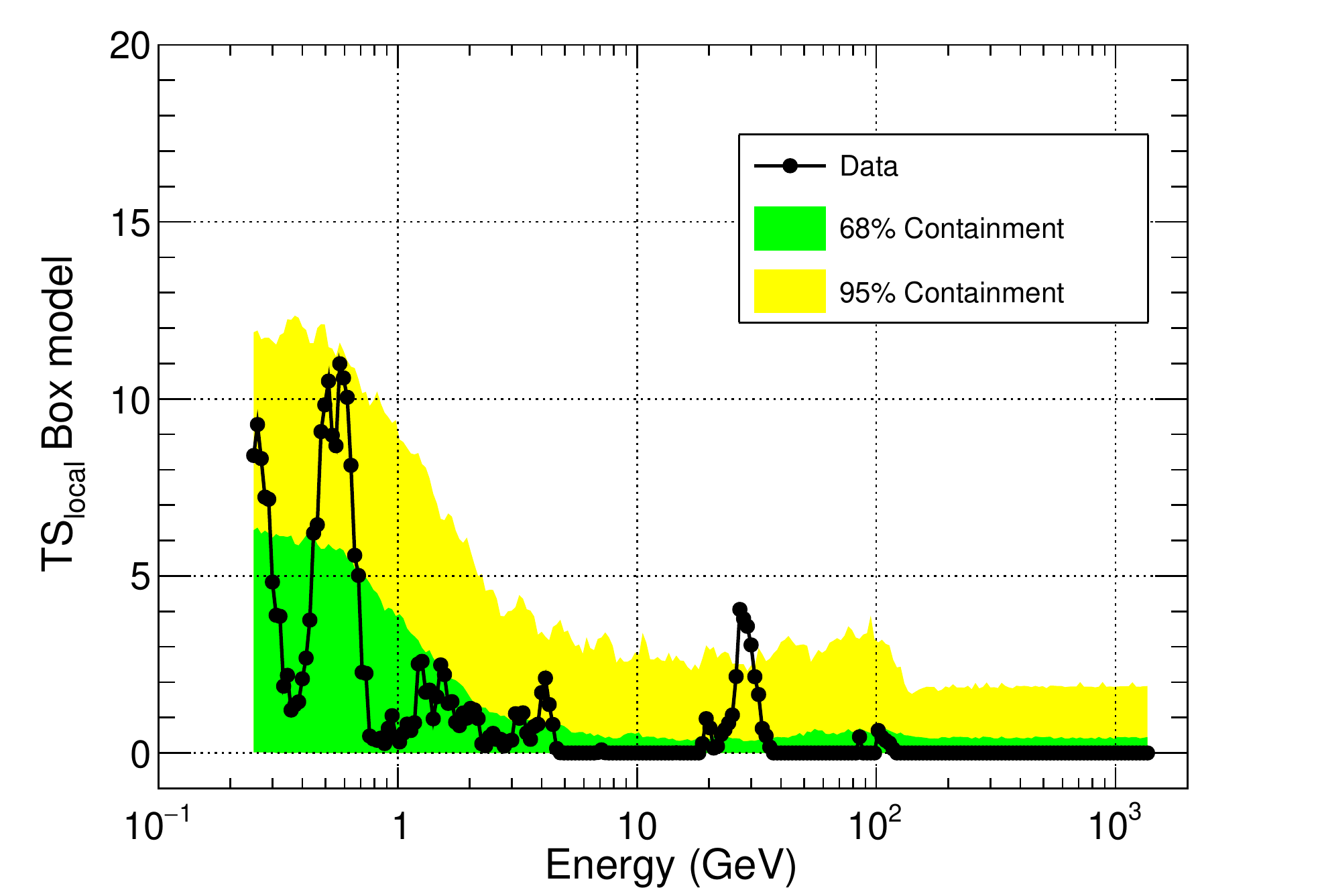}
\includegraphics[width=\columnwidth,height=0.25\textheight,clip]{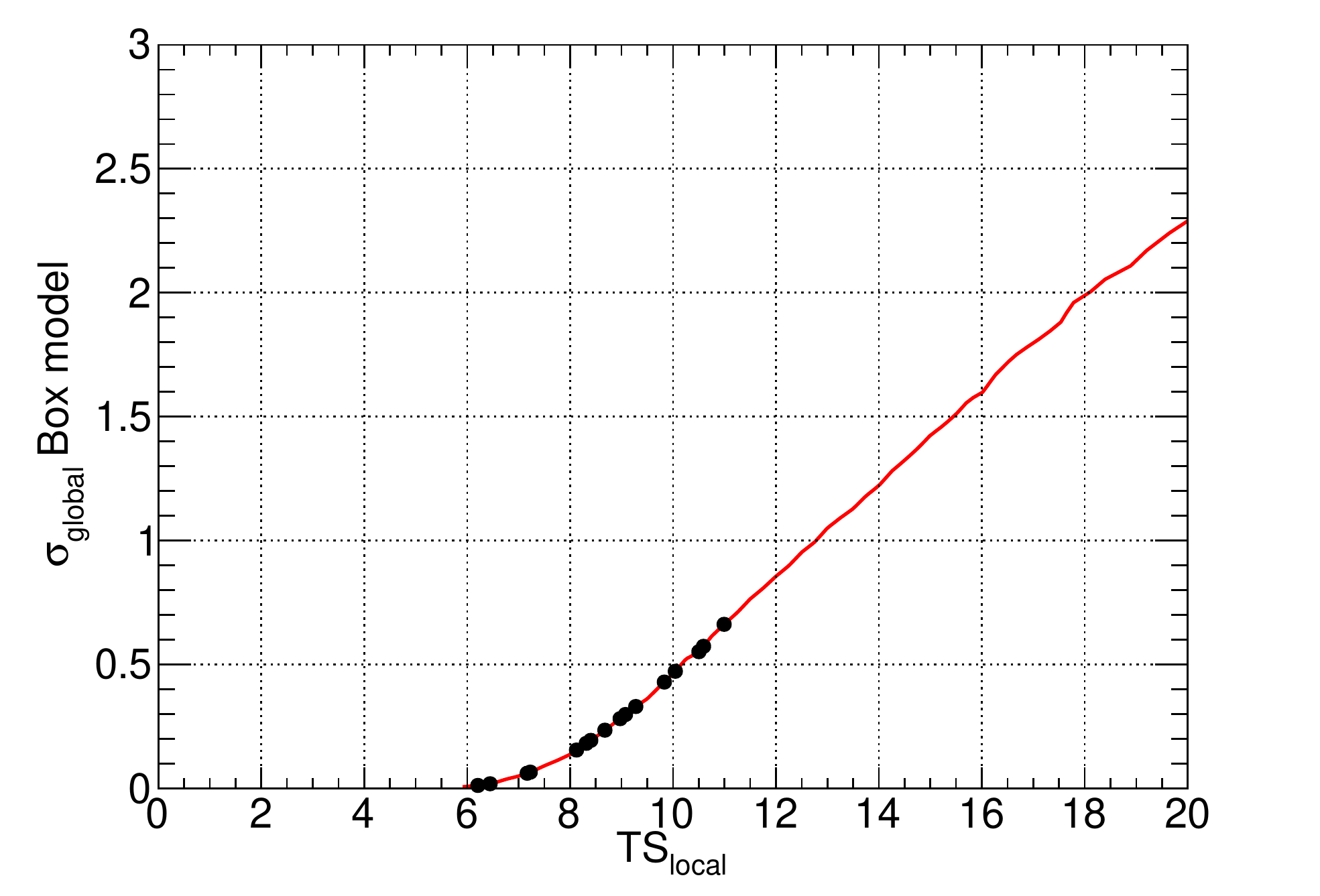}
\includegraphics[width=\columnwidth,height=0.25\textheight,clip]{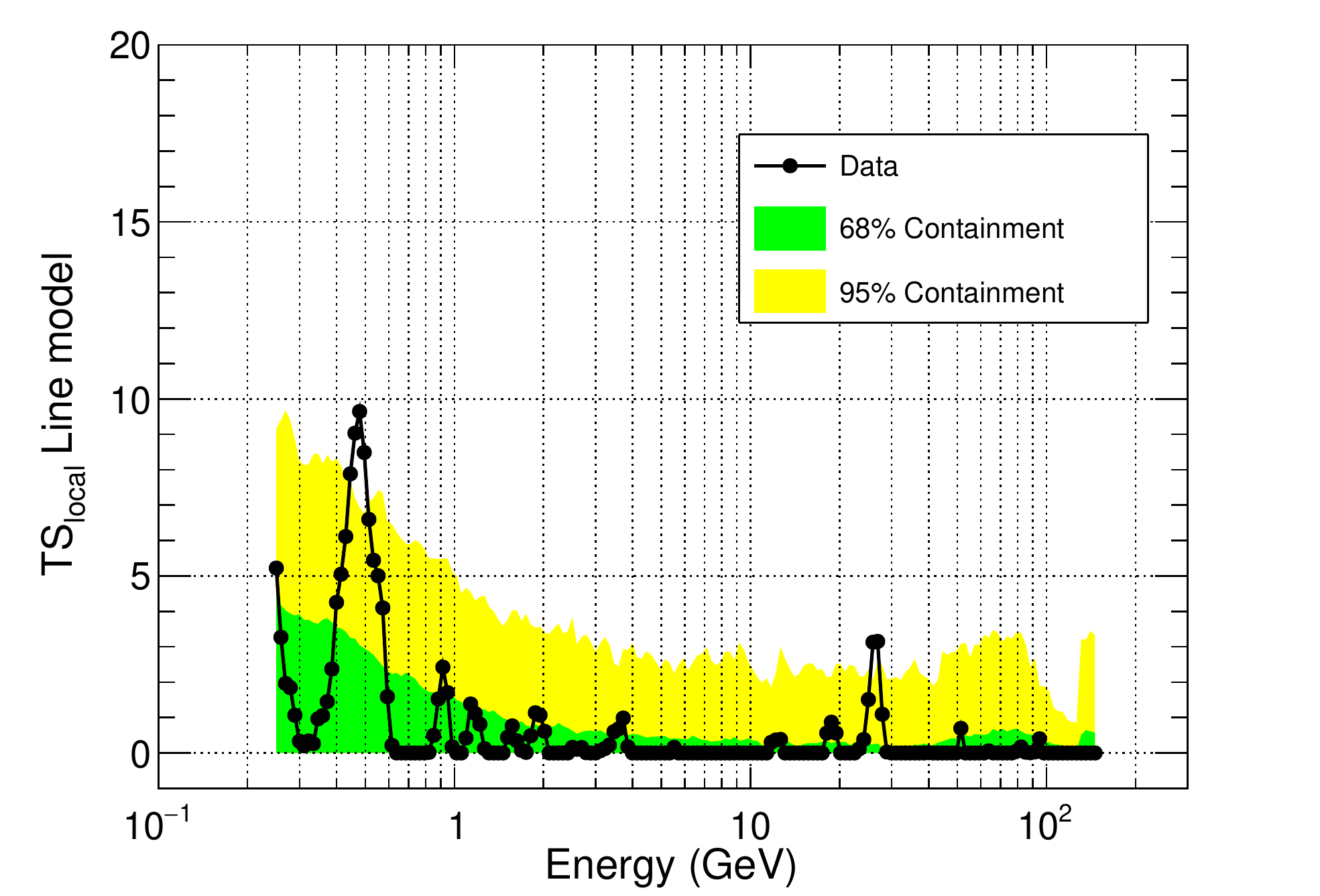}
\includegraphics[width=\columnwidth,height=0.25\textheight,clip]{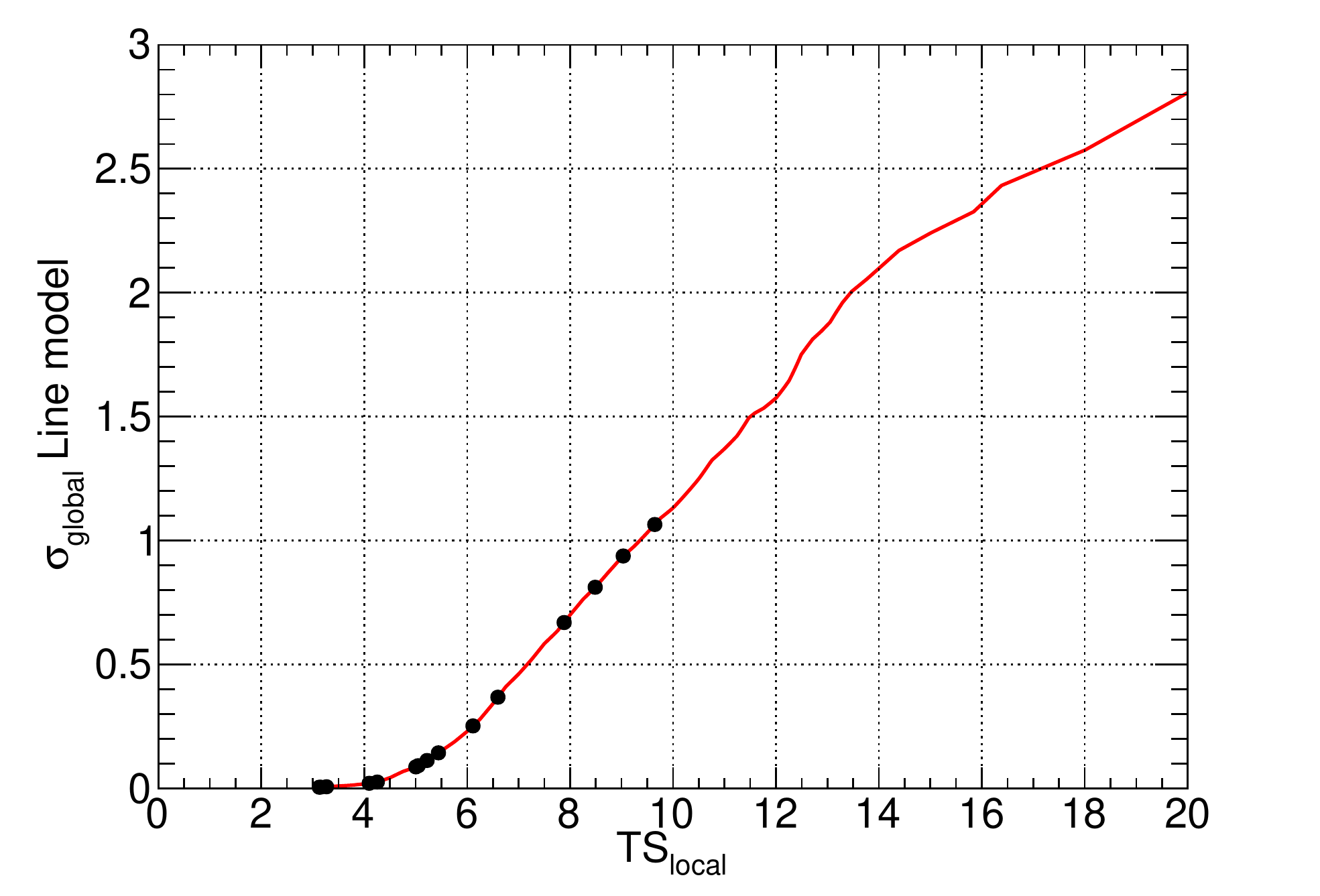}
\caption{Local TS (left plots) and global significance (right plots) of the possible features in the solar gamma-ray spectrum. Top plots: box model; bottom plots: line model. 
The green and yellow regions in the left plots indicate the one-sided 68\% and 95\% expectation bands for the TS evaluated from the pseudo-experiments.
The red lines in the right plots indicate the conversion from $TS_{\text{local}}$ to $\sigma_{\text{global}}$, while the black dots correspond the most significant features. }
\label{fig:ts}
\end{figure*}

As discussed above, we perform the fits in the observed energy range from $100\units{MeV}$ up to $150\units{GeV}$, with the last window extending from $37.5 \units{GeV}$ to $150\units{GeV}$ and centered at about $90\units{GeV}$. In the case of the line, the upper bound $E_{\text{f,max}}$ for the characteristic energy of the feature to be fitted has been chosen by requiring that at least half of the photons from the line have observed energies in the last fit window, and it is $E_{\text{f,max}}=150\units{GeV}$. In the case of the box, we have chosen $E_{\text{f,max}}=1.8\units{TeV}$, which corresponds to the maximum energy where the Instrument Response Functions (IRFs) have been evaluated~\cite{Bruel:2018lac}.

Figure~\ref{fig:ul} shows respectively the ULs at $95\%$ confidence level on the intensity of the box-like feature (left panel) and of the line-like feature feature (right panel), respectively. The plots also show the central 68\% and 95\% containment bands for the ULs, evaluated using the pseudo-experiment technique described in Sec.~\ref{sec:ana}
\footnote{For each energy window we have built the distribution of the ULs derived from the pseudo-experiments and we have evaluated its quantiles. The 68\% (95\%) central containment bands correspond to the intervals between the 16\% and 84\% (2.5\% and 97.5\%) quantiles of these distributions.}. 
For both models, the fitted parameters lie within the central $95\%$ confidence belt in all the energy windows. In the case of the box-like feature, the limit on the intensity becomes constant above $150\units{GeV}$ because all possible box-like features with characteristic energy above this value will yield a roughly uniform photon spectrum in the last fit window. In the case of the line-like feature, the limit on the line intensity above $\sim 100\units{GeV}$ increases with the characteristic energy of the line, as larger fractions of photons with observed energies above $150\units{GeV}$ are predicted by the model.

Figure~\ref{fig:ts} shows the local significance (left plots) and the global significance (right plots) of the possible features at different energies for both the models considered in this work. In the left plots we also show the one-sided 68\% and 95\% expectation bands for the TS evaluated from the pseudo-experiments. We note that the $TS_{\rm local}$ values exhibit a peak in the energy windows around $E=500\units{MeV}$ for both the box and line features. This is due to a change in the slope of the observed count distribution (see Fig.\ref{fig:counts}) that is interpreted by our fitting algorithm as a ``feature'' in the spectrum. Nevertheless, in both cases, we see that in nearly all energy windows the values of $TS_{\rm local}$ lie within the $95\%$ expectation band, although there are a few values of $TS_{\rm local}$ which are slightly above the expectation band. 
All these possible features, which appear to be locally significant, turn out to be globally insignificant. In fact, in the evaluation of the global significance, the fact that the fits are not independent should be considered, and the number of trials performed should be taken into account. To convert the $TS_{\rm local}$ into a global significance we have used the results of the 1000 pseudo-experiments\footnote{Each pseudo experiment corresponds to a simulation of one full search across the whole energy range.} discussed in Sec.~\ref{sec:ana}. For each pseudo-experiment, we record the largest value of the local Test Statistic $TS_{\rm max}$; then, we evaluate the quantiles of the distribution of $TS_{\rm max}$, and the corresponding values of the global significance $\sigma_{\rm global}$, assuming that it obeys to a half-normal distribution.
The results after the global significance conversion are shown in the right-hand plots in Fig.~\ref{fig:ts} for the box-like feature (top right panel) and for the line-like feature (bottom right panel). The most significant features have global significances less than or comparable to 1$\sigma$ in both cases, and therefore are globally insignificant.

In our analysis procedure we have assumed that a possible background feature appears with the same amplitude in both the ``on'' and ``off'' regions. However, 
since we observe a different number of counts in the ``on'' and ``off'' regions (see Fig.~\ref{fig:counts}), the background feature could manifest with different amplitudes in the two regions. In this case it could be reasonable to assume that the intensity of the background feature scales with the observed events. We have therefore implemented an alternative approach in our fitting procedure, in which the parameter $s_{\text{bkg}}$ of the background feature in the ``on'' region is scaled by a factor $N_{\text{on}}/N_{\text{off}}$, where  $N_{\text{on}}$ and $N_{\text{off}}$ are the number of observed gamma rays in the ``on'' and ``off'' regions and in the energy window considered in the fit, i.e. in the ``on'' region $s_{\text{bkg}}$ is replaced with $s_{\text{bkg}} N_{\text{on}}/N_{\text{off}}$. We find that with this approach 
the TS peaks become reduced further 
and the upper limits do not change with respect to those shown in Fig.~\ref{fig:ul}.

The limits on the intensity of the features can be converted into limits on the flux at Earth $\Phi_{\text{DM}}(E)$ of gamma rays from solar DM. In turn, using Eq.\ref{eq:phidm} for the mediator scenario, 
the flux limits can be converted into limits on the DM-nucleon cross section.
We point out here that the DM gamma-ray flux at Earth depends on the DM capture rate $\Gamma_{\text{cap}}$ in the Sun, which in turn depends on the DM scenario considered. As discussed in Sec.\ref{sec:models}, in the case of DM  annihilation via a long-lived intermediate state, the capture rate calculation has been performed using the {\tt DARKSUSY} code. On the other hand, for the inelastic scattering scenario, we are not able to perform the calculation of the capture rate, since at present all available models in the literature start from energies of about 100\units{GeV}, and are therefore valid above the energy range explored in our analysis \cite{Menon:2009qj}. 

In the mediator scenario,  using the values of $\Phi_{\text{DM}}(E)$ calculated with the {\tt DARKSUSY} code (Fig.~\ref{fig:dmflux}), we have evaluated the limits on the DM-nucleon cross section as:

\begin{equation}
    \sigma_{\text{UL}}(m_\chi) = \frac{\Phi_{\text{UL}}(E=m_\chi)}{\Phi_{\text{DM}}(E=m_\chi)} \times 10^{-40}~\units{cm^2}.
    \label{eq:sigmalim}
\end{equation}
The value of $10^{-40} \units{cm^2}$ in the previous equation corresponds to the value of the DM-nucleon cross section used to calculate the capture rates shown in Figure~\ref{fig:dmflux}. 

\begin{figure}[!ht]
    \centering
    \includegraphics[width=\columnwidth,height=0.25\textheight,clip]{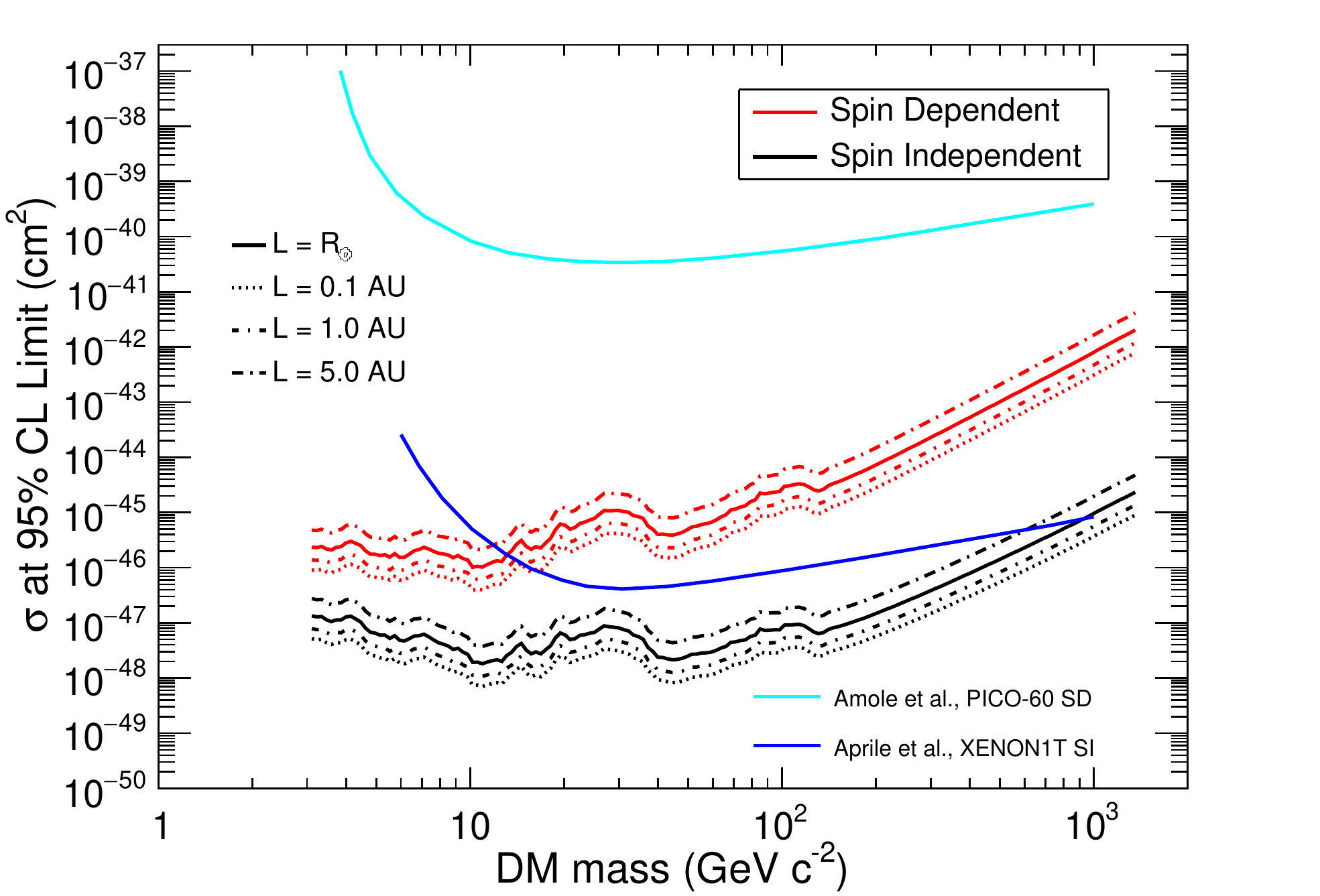}
    \caption{Limits at 95\%CL on the DM-nucleon cross section via long-lived mediator for four different decay lengths $L=R_{\odot}$, $0.1$, $1$ and $5\units{AU}$. 
    The plots also show the limits at $90\%$ CL from the PICO-60 experiment~\cite{Amole:2019fdf} in the case of spin-dependent (SD) scattering (cyan line) and from the XENON1T experiment~\cite{Aprile:2018dbl} in the case of spin-independent (SI) scattering (blue line).}
    \label{fig:ulsigma}
    \end{figure}

Figure~\ref{fig:ulsigma} shows the constraints on DM annihilation to $4\gamma$ via an intermediate state, obtained from the upper limits on the intensity of the box feature, assuming that DM capture takes place either via spin-independent scattering (black lines) or spin-dependent scattering (red lines). The constraints have been calculated for four values of the decay length of the intermediate state, $L=R_{\odot}$, $0.1$, $1$ and $5\units{AU}$. The limits on the cross-sections evaluated in this work for the DM mass from $3\units{GeV}$ up to $150\units{GeV/c^2}$ are in the range from about $10^{-46}$ to $10^{-45}\units{cm^{2}}$ for the spin-dependent cross section and in the range from about $10^{-48}$ to $10^{-47}\units{cm^{2}}$ for the spin-independent cross section. The ranges of variation depend on the life time of the mediator considered. We consider DM masses above $3\units{GeV}$ since, as mentioned in Sec.~\ref{sec:models}, we do not take the evaporation mechanism into account.
Above $150\units{GeV/c^2}$ the ULs on the intensity of the box features are constant, while the capture rate decreases as $m_{\chi}^{-2}$. For this reason the limits on the cross section increase with the DM mass. 
In Figure~\ref{fig:ulsigma} we also compare our limits with those obtained from direct measurements of the spin-independent and of the spin-dependent DM-nucleon cross sections performed by the XENON1T~\cite{Aprile:2018dbl} and PICO-60~\cite{Amole:2019fdf} experiments respectively. In the case of the spin-dependent cross section our limits are a few orders of magnitude stronger than those obtained by PICO-60 in the whole DM mass range explored, while in the case of the spin-independent cross section are consistent with those from XENON1T.\footnote{We remark here that the constraints quoted in Refs.~\cite{Aprile:2018dbl} and ~\cite{Amole:2019fdf} are upper bounds at 90\% CL, while here we are presenting 95\% CL limits.}

\begin{figure}[!ht]
    \centering
    \includegraphics[width=\columnwidth,height=0.25\textheight,clip]{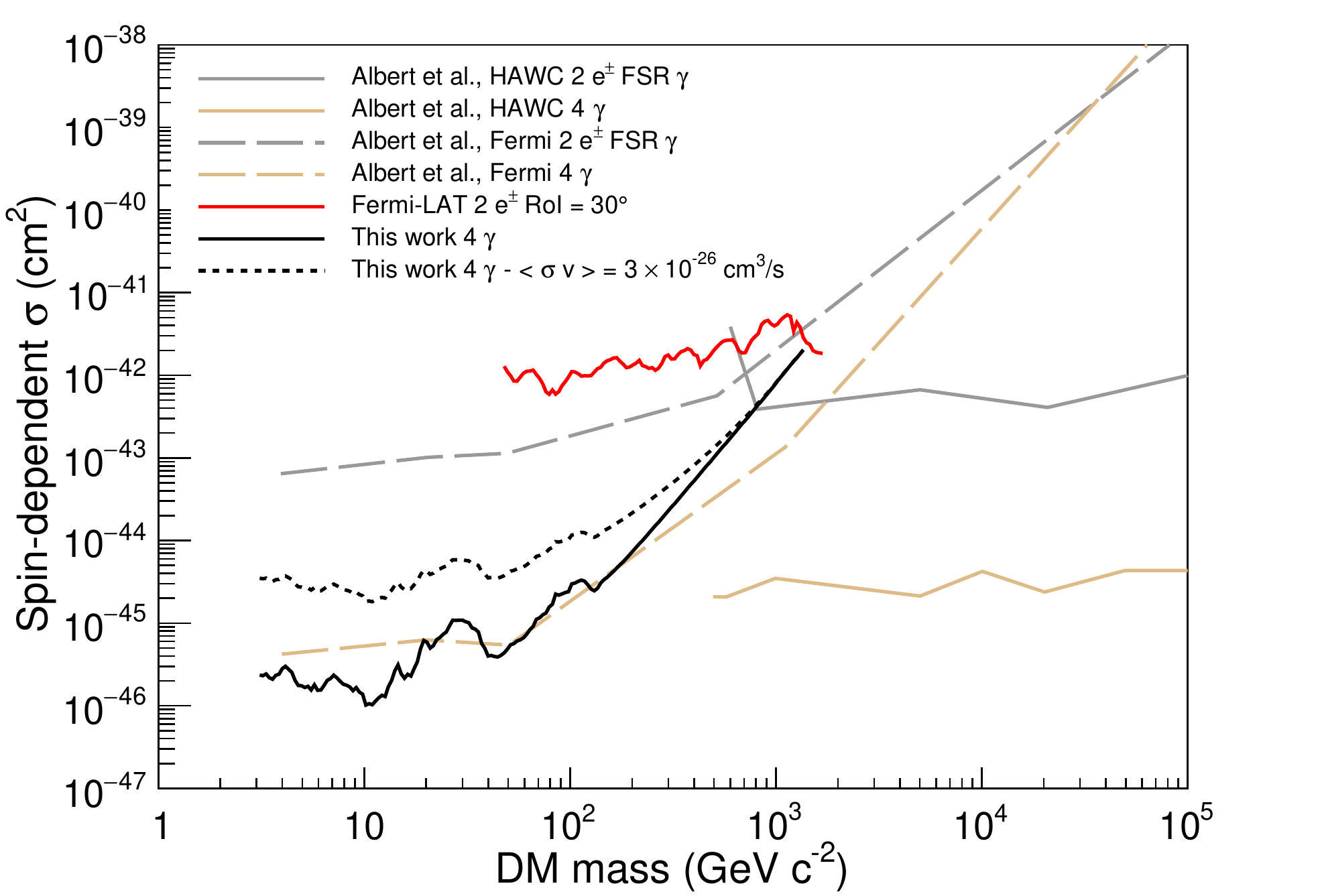}
    \caption{Upper limits at 95\%CL on the spin-dependent DM-nucleon scattering cross section for the long lived mediator with decay length $L=R_{\odot}$. Our results (black line) are shown together with those from the HAWC and Fermi~\cite{Albert:2018jwh} with gamma rays from DM annihilating into 2 $e^{\pm}$, i.e., 2 $e^{\pm}$ FSR $\gamma$ case (gray lines) and into 4 $\gamma$ (brown lines), and electron/positron Fermi-LAT results (red line)~\cite{Mazziotta:2019CREs}. Our results corrected for the non equilibrium state with $\langle \sigma_{ann} v \rangle =3 \times 10^{-26} \units{cm^3/s}$ is shown with the dotted line.}
    \label{fig:expcomp}
\end{figure}

\section{Discussion and Conclusion}
\label{sec:con}

In this work we have implemented a dedicated analysis based on a Poisson maximum likelihood fitting approach in order to search for potential dark matter features in the energy spectrum of gamma rays coming from the Sun using a 10-year Fermi-LAT dataset. We have considered two possible scenarios in which gamma rays are produced from  DM particles captured in the Sun that annihilate either directly or via a long-lived mediator stage.

Starting from the limits on the strength of the features and calculating the expected flux at Earth of gamma rays, we have constrained the DM-nucleon cross sections. The DM gamma-ray flux  at Earth depends on the capture rate $\Gamma_{\text{cap}}$, which, in the case of the mediator scenario, has been calculated using the {\tt DARKSUSY} code.
In the case of the line-like feature, we have not converted the upper limit on the line intensities into DM-nucleon cross section constraints  because there are no models presently available in the literature for the calculation of the capture rate. Nonetheless, the limits obtained in our analysis can be used, in principle, to study constraints on the DM models involving a line-like feature as DM signature.
In case of the long-lived mediator scenario, the limits on the spin-independent cross-section are in the range from about $10^{-48}\units{cm^2}$ to about $10^{-47}\units{cm^2}$, while those on the spin-dependent cross section are in the range from about $10^{-46}\units{cm^2}$ to about $10^{-45}\units{cm^2}$ for DM masses between $3\units{GeV/c^2}$ and $150\units{GeV/c^2}$.
These limits depend on the decay length of the mediator, and we have considered different decay lengths in the range from $R_{\odot}$ to $5 \units{AU}$. 
For DM masses up to $1.8 \units{TeV/c^2}$ the limits increase with the mass, due to the dependence of the capture rate on the DM mass.
Above this mass a further increase of the limits is expected. However, as mentioned above, we limit our analysis up to $1.8\units{TeV/c^2}$, corresponding to the maximum energy for which the IRFs have been evaluated.

The model considered in this work is the same investigated by other authors in their recent works~\cite{Albert:2018jwh,Leane:2017vag}. 
A summary of the limits on the spin-dependent DM-nucleon cross section for the long-lived mediator scenario with decay length $L=R_{\odot}$ is shown in Fig.~\ref{fig:expcomp}, together with the recent constraints obtained from the analysis of solar gamma rays by HAWC and Fermi~\cite{Albert:2018jwh} for the case discussed in this work, in which the mediator decays into 2 photons (4$\gamma$ case), and for the case in which the mediator decays into $e^{\pm}$ pairs and photons are produced as final state radiation (2 $e^{\pm}$ FSR $\gamma$ case)~\cite{Mazziotta:2019CREs}.
The results obtained in this work are consistent with those from HAWC and Fermi in~\cite{Albert:2018jwh}, in which the limits on the spin-dependent cross section are evaluated for dark matter masses between 4\units{GeV/c^2} and $10^6\units{GeV/c^2}$ combining Fermi and HAWC 3-years observations of the Sun. However, we point out here that the authors of Ref.~\cite{Albert:2018jwh} used a different strategy to compute the constraints on the spin-dependent DM-nucleon cross section. 
In fact, for each value of the DM mass $m_{DM}$, they evaluate the upper limit on $\sigma$ imposing that the DM-induced signal must not exceed the flux allowed by the Fermi-LAT measurement and by the HAWC sensitivity in any energy bin.

Also relevant is to point out that the analysis of Ref.~\cite{Albert:2018jwh} was performed using a data set taken during a period of maximum solar activity, when the solar steady gamma-ray emission reaches its minimum. Since the intensity of a DM-induced signal feature in the solar gamma-ray spectrum should be independent of the solar activity, a search of a feature limited to the period of maximum solar activity should benefit from an higher signal-to-noise ratio. We have therefore repeated our analysis with a restricted data sample, corresponding to the period from January 2011 to December 2015, where the solar activity was at its maximum. 
The upper limits on the intensities of possible features are actually slightly worse on average by $10-20\%$ above $2 \units{GeV}$ with respect to those obtained when analyzing the full data sample, while below $2 \units{GeV}$ they are compatible, although showing relative fluctuations of order $30\%$. We thus conclude that there is no significant gain in considering the solar maximum data period.

As mentioned in section~\ref{sec:models}, our results, as well as those in~\cite{Albert:2018jwh,Leane:2017vag}, have been derived assuming equilibrium between DM capture and annihilation in the Sun. This condition is met only if the equilibrium time scale $\tau$ is less than the age of the Sun. In Refs.~\cite{Jungman:1995df, Griest:1986yu} it is shown that, assuming for the lifetime of the Sun the commonly accepted value $t=4.5\units{Gyr}$, the ratio $t/\tau$ is given by:

\begin{equation}
\label{eq:tovertau}
\frac{t}{\tau} = 330 \left( \frac{\Gamma_{cap}}{\units{s^{-1}}} \right)^{1/2} \left( \frac{\langle \sigma_{ann} v \rangle}{\units{cm^{3}/s}}\right)^{1/2} \left( \frac{m_{\chi}}{10\units{GeV}} \right)^{3/4}
\end{equation}

If equilibrium is not reached, the annihilation rate is reduced of a factor $\tanh^2(t/\tau)$, and consequently the limits on the DM-nucleon cross sections must be evaluated taking this factor into account in eqs.~\ref{eq:caprate} and~\ref{eq:sigmalim}. 
Starting from the limits obtained in the equilibrium scenario (Figure~\ref{fig:ulsigma}), we have implemented a numerical procedure to evaluate limits in non-equilibrium scenarios, in the case of spin-dependent cross section. In Figure ~\ref{fig:expcomp} the limits in the equilibrium setup are shown together with those in a non-equilibrium setup. For the non-equilibrium setup, we have assumed a velocity averaged cross section at the level of the thermal relic value $\langle \sigma_{ann} v \rangle =3 \times 10^{-26} \units{cm^3/s}$~\cite{Steigman:2012nb} (for current limits on this total DM annihilation cross section, see, e.g.,~\cite{Ackermann:2015lka,Depta:2019lbe,Jedamzik:2009uy,Slatyer:2015jla}). However, the relevant annihilation cross section to induce our specific (line-like or box-shaped) gamma-ray signal is significantly more strongly constrained, see e.g. refs.~\cite{Ibarra:2013eda,Ackermann:2015lka,Li:2018rqo,Ibarra:2012dw,Jedamzik:2009uy,Depta:2019lbe,Slatyer:2015jla}. The general procedure to rescale our limits to any DM setup is therefore provided by Eqs.~\ref{eq:eq12} and~\ref{eq:eq13}, as described below.
We see that for DM masses below $100\units{GeV}$ the non-equilibrium constraints are a factor $\sim 10$ weaker than those in the equilibrium scenario, while for DM masses in the $\units{TeV}$ range the two scenarios yield similar limits. 

Using the capture rates shown in Figure~\ref{fig:dmflux} we have found that, if $\langle \sigma_{ann} v \rangle $ is of the order of the thermal relic cross section, a DM-nucleon cross section of the order of $10^{-43} \units{cm^2}$ ($10^{-45} \units{cm^2}$) in the spin-dependent (spin-independent) case is needed to attain equilibrium in the DM mass range explored in the present analysis. Therefore  any constraints on $\sigma$ below these values, like those presented in Ref.~\cite{Albert:2018jwh}, should be corrected to take into account the non-equilibrium between the annihilation and capture processes.

Finally, we point out that the limits in the elastic scattering scenario are evaluated assuming $100\%$ branching ratio for DM annihilating into a pair of mediators $\phi$s, which in turn decay into a pair of gamma rays. If there are additional channels in the DM particle annihilations and in the the mediator decay, the branching ratios should be properly taken into account. In this case, the dark matter flux in eq.~\ref{eq:phidm} should be scaled by a factor $\mathcal{BR}$, where $\mathcal{BR}$ is the product of the branching ratios of the processes $\chi \chi \rightarrow \phi \phi$ and $\phi \rightarrow \gamma \gamma$, and therefore the upper limit on the DM-nucleon cross section in eq.~\ref{eq:sigmalim} would be scaled by a factor $1/\mathcal{BR}$.

More precisely, to account for non-equilibrium and for the $\mathcal{BR}$, the DM-nucleon cross section limit $\sigma$ at a given DM mass $m_\chi$ should be evaluated by solving the following equation:

\begin{equation}
    \mathcal{BR} ~ \sigma \tanh^2 \left( f \sqrt{\langle \sigma_{ann} v \rangle} \sqrt{\sigma} \right) = \sigma_{eq}
    \label{eq:eq12}
\end{equation}
where $\sigma_{eq}$ is the DM-nucleon cross section limit evaluated in the case of equilibrium. Following Eq.~\ref{eq:tovertau}, the factor $f$ is given by:
\begin{equation}
f = 330 \left( \frac{\Gamma_{cap,ref}}{\units{s^{-1}}} \right)^{1/2} 
\left(  \frac{\sigma_{ref}}{cm^{2}}     \right)^{-1/2}
\left(  \frac{m_{\chi}}{10\units{GeV}} \right)^{3/4}
\label{eq:eq13}
\end{equation}
where $\sigma_{ref}=10^{-40}\units{cm^{2}}$ is the cross section used to calculate the reference values of the capture rates $\Gamma_{cap,ref}$, as discussed in Sec.~\ref{sec:inter}.

\begin{acknowledgments} 
The Fermi LAT Collaboration acknowledges generous ongoing support
from a number of agencies and institutes that have supported both the
development and the operation of the LAT as well as scientific data analysis.
These include the National Aeronautics and Space Administration and the
Department of Energy in the United States, the Commissariat \`a l'Energie Atomique
and the Centre National de la Recherche Scientifique / Institut National de Physique
Nucl\'eaire et de Physique des Particules in France, the Agenzia Spaziale Italiana
and the Istituto Nazionale di Fisica Nucleare in Italy, the Ministry of Education,
Culture, Sports, Science and Technology (MEXT), High Energy Accelerator Research
Organization (KEK) and Japan Aerospace Exploration Agency (JAXA) in Japan, and
the K.~A.~Wallenberg Foundation, the Swedish Research Council and the
Swedish National Space Board in Sweden.
 
Additional support for science analysis during the operations phase is gratefully
acknowledged from the Istituto Nazionale di Astrofisica in Italy and the Centre
National d'\'Etudes Spatiales in France. This work performed in part under DOE
Contract DE-AC02-76SF00515.
\end{acknowledgments}

\bibliographystyle{apsrev4-1}
\bibliography{GammaRaySunDM.bib}{}

\begin{thebibliography}{51}%
\makeatletter
\providecommand \@ifxundefined [1]{%
 \@ifx{#1\undefined}
}%
\providecommand \@ifnum [1]{%
 \ifnum #1\expandafter \@firstoftwo
 \else \expandafter \@secondoftwo
 \fi
}%
\providecommand \@ifx [1]{%
 \ifx #1\expandafter \@firstoftwo
 \else \expandafter \@secondoftwo
 \fi
}%
\providecommand \natexlab [1]{#1}%
\providecommand \enquote  [1]{``#1''}%
\providecommand \bibnamefont  [1]{#1}%
\providecommand \bibfnamefont [1]{#1}%
\providecommand \citenamefont [1]{#1}%
\providecommand \href@noop [0]{\@secondoftwo}%
\providecommand \href [0]{\begingroup \@sanitize@url \@href}%
\providecommand \@href[1]{\@@startlink{#1}\@@href}%
\providecommand \@@href[1]{\endgroup#1\@@endlink}%
\providecommand \@sanitize@url [0]{\catcode `\\12\catcode `\$12\catcode
  `\&12\catcode `\#12\catcode `\^12\catcode `\_12\catcode `\%12\relax}%
\providecommand \@@startlink[1]{}%
\providecommand \@@endlink[0]{}%
\providecommand \url  [0]{\begingroup\@sanitize@url \@url }%
\providecommand \@url [1]{\endgroup\@href {#1}{\urlprefix }}%
\providecommand \urlprefix  [0]{URL }%
\providecommand \Eprint [0]{\href }%
\providecommand \doibase [0]{http://dx.doi.org/}%
\providecommand \selectlanguage [0]{\@gobble}%
\providecommand \bibinfo  [0]{\@secondoftwo}%
\providecommand \bibfield  [0]{\@secondoftwo}%
\providecommand \translation [1]{[#1]}%
\providecommand \BibitemOpen [0]{}%
\providecommand \bibitemStop [0]{}%
\providecommand \bibitemNoStop [0]{.\EOS\space}%
\providecommand \EOS [0]{\spacefactor3000\relax}%
\providecommand \BibitemShut  [1]{\csname bibitem#1\endcsname}%
\let\auto@bib@innerbib\@empty
\bibitem [{\citenamefont {Tanabashi}\ \emph {et~al.}(2018)\citenamefont
  {Tanabashi} \emph {et~al.}}]{PhysRevD.98.030001}%
  \BibitemOpen
  \bibfield  {author} {\bibinfo {author} {\bibfnamefont {M.}~\bibnamefont
  {Tanabashi}} \emph {et~al.} (\bibinfo {collaboration} {Particle Data
  Group}),\ }\href {\doibase 10.1103/PhysRevD.98.030001} {\bibfield  {journal}
  {\bibinfo  {journal} {Phys. Rev. D}\ }\textbf {\bibinfo {volume} {98}},\
  \bibinfo {pages} {030001} (\bibinfo {year} {2018})}\BibitemShut {NoStop}%
\bibitem [{\citenamefont {Pospelov}\ \emph {et~al.}(2008)\citenamefont
  {Pospelov}, \citenamefont {Ritz},\ and\ \citenamefont
  {Voloshin}}]{Pospelov:2007mp}%
  \BibitemOpen
  \bibfield  {author} {\bibinfo {author} {\bibfnamefont {M.}~\bibnamefont
  {Pospelov}}, \bibinfo {author} {\bibfnamefont {A.}~\bibnamefont {Ritz}}, \
  and\ \bibinfo {author} {\bibfnamefont {M.~B.}\ \bibnamefont {Voloshin}},\
  }\href {\doibase 10.1016/j.physletb.2008.02.052} {\bibfield  {journal}
  {\bibinfo  {journal} {Phys. Lett.}\ }\textbf {\bibinfo {volume} {B662}},\
  \bibinfo {pages} {53} (\bibinfo {year} {2008})},\ \Eprint
  {http://arxiv.org/abs/0711.4866} {arXiv:0711.4866 [hep-ph]} \BibitemShut
  {NoStop}%
\bibitem [{\citenamefont {Arkani-Hamed}\ \emph {et~al.}(2009)\citenamefont
  {Arkani-Hamed}, \citenamefont {Finkbeiner}, \citenamefont {Slatyer},\ and\
  \citenamefont {Weiner}}]{ArkaniHamed:2008qn}%
  \BibitemOpen
  \bibfield  {author} {\bibinfo {author} {\bibfnamefont {N.}~\bibnamefont
  {Arkani-Hamed}}, \bibinfo {author} {\bibfnamefont {D.~P.}\ \bibnamefont
  {Finkbeiner}}, \bibinfo {author} {\bibfnamefont {T.~R.}\ \bibnamefont
  {Slatyer}}, \ and\ \bibinfo {author} {\bibfnamefont {N.}~\bibnamefont
  {Weiner}},\ }\href {\doibase 10.1103/PhysRevD.79.015014} {\bibfield
  {journal} {\bibinfo  {journal} {Phys. Rev.}\ }\textbf {\bibinfo {volume}
  {D79}},\ \bibinfo {pages} {015014} (\bibinfo {year} {2009})},\ \Eprint
  {http://arxiv.org/abs/0810.0713} {arXiv:0810.0713 [hep-ph]} \BibitemShut
  {NoStop}%
\bibitem [{\citenamefont {Schuster}\ \emph
  {et~al.}(2010{\natexlab{a}})\citenamefont {Schuster}, \citenamefont {Toro},
  \citenamefont {Weiner},\ and\ \citenamefont {Yavin}}]{Schuster:2009fc}%
  \BibitemOpen
  \bibfield  {author} {\bibinfo {author} {\bibfnamefont {P.}~\bibnamefont
  {Schuster}}, \bibinfo {author} {\bibfnamefont {N.}~\bibnamefont {Toro}},
  \bibinfo {author} {\bibfnamefont {N.}~\bibnamefont {Weiner}}, \ and\ \bibinfo
  {author} {\bibfnamefont {I.}~\bibnamefont {Yavin}},\ }\href {\doibase
  10.1103/PhysRevD.82.115012} {\bibfield  {journal} {\bibinfo  {journal} {Phys.
  Rev.}\ }\textbf {\bibinfo {volume} {D82}},\ \bibinfo {pages} {115012}
  (\bibinfo {year} {2010}{\natexlab{a}})},\ \Eprint
  {http://arxiv.org/abs/0910.1839} {arXiv:0910.1839 [hep-ph]} \BibitemShut
  {NoStop}%
\bibitem [{\citenamefont {Leane}\ \emph {et~al.}(2017)\citenamefont {Leane},
  \citenamefont {Ng},\ and\ \citenamefont {Beacom}}]{Leane:2017vag}%
  \BibitemOpen
  \bibfield  {author} {\bibinfo {author} {\bibfnamefont {R.~K.}\ \bibnamefont
  {Leane}}, \bibinfo {author} {\bibfnamefont {K.~C.~Y.}\ \bibnamefont {Ng}}, \
  and\ \bibinfo {author} {\bibfnamefont {J.~F.}\ \bibnamefont {Beacom}},\
  }\href {\doibase 10.1103/PhysRevD.95.123016} {\bibfield  {journal} {\bibinfo
  {journal} {Phys. Rev.}\ }\textbf {\bibinfo {volume} {D95}},\ \bibinfo {pages}
  {123016} (\bibinfo {year} {2017})},\ \Eprint
  {http://arxiv.org/abs/1703.04629} {arXiv:1703.04629 [astro-ph.HE]}
  \BibitemShut {NoStop}%
\bibitem [{\citenamefont {Schuster}\ \emph
  {et~al.}(2010{\natexlab{b}})\citenamefont {Schuster}, \citenamefont {Toro},\
  and\ \citenamefont {Yavin}}]{Schuster:2009au}%
  \BibitemOpen
  \bibfield  {author} {\bibinfo {author} {\bibfnamefont {P.}~\bibnamefont
  {Schuster}}, \bibinfo {author} {\bibfnamefont {N.}~\bibnamefont {Toro}}, \
  and\ \bibinfo {author} {\bibfnamefont {I.}~\bibnamefont {Yavin}},\ }\href
  {\doibase 10.1103/PhysRevD.81.016002} {\bibfield  {journal} {\bibinfo
  {journal} {Phys. Rev.}\ }\textbf {\bibinfo {volume} {D81}},\ \bibinfo {pages}
  {016002} (\bibinfo {year} {2010}{\natexlab{b}})},\ \Eprint
  {http://arxiv.org/abs/0910.1602} {arXiv:0910.1602 [hep-ph]} \BibitemShut
  {NoStop}%
\bibitem [{\citenamefont {Bell}\ and\ \citenamefont
  {Petraki}(2011)}]{Bell:2011sn}%
  \BibitemOpen
  \bibfield  {author} {\bibinfo {author} {\bibfnamefont {N.~F.}\ \bibnamefont
  {Bell}}\ and\ \bibinfo {author} {\bibfnamefont {K.}~\bibnamefont {Petraki}},\
  }\href {\doibase 10.1088/1475-7516/2011/04/003} {\bibfield  {journal}
  {\bibinfo  {journal} {JCAP}\ }\textbf {\bibinfo {volume} {1104}},\ \bibinfo
  {pages} {003} (\bibinfo {year} {2011})},\ \Eprint
  {http://arxiv.org/abs/1102.2958} {arXiv:1102.2958 [hep-ph]} \BibitemShut
  {NoStop}%
\bibitem [{\citenamefont {Arina}\ \emph {et~al.}(2017)\citenamefont {Arina},
  \citenamefont {Backovic}, \citenamefont {Heisig},\ and\ \citenamefont
  {Lucente}}]{Arina:2017sng}%
  \BibitemOpen
  \bibfield  {author} {\bibinfo {author} {\bibfnamefont {C.}~\bibnamefont
  {Arina}}, \bibinfo {author} {\bibfnamefont {M.}~\bibnamefont {Backovic}},
  \bibinfo {author} {\bibfnamefont {J.}~\bibnamefont {Heisig}}, \ and\ \bibinfo
  {author} {\bibfnamefont {M.}~\bibnamefont {Lucente}},\ }\href {\doibase
  10.1103/PhysRevD.96.063010} {\bibfield  {journal} {\bibinfo  {journal} {Phys.
  Rev.}\ }\textbf {\bibinfo {volume} {D96}},\ \bibinfo {pages} {063010}
  (\bibinfo {year} {2017})},\ \Eprint {http://arxiv.org/abs/1703.08087}
  {arXiv:1703.08087 [astro-ph.HE]} \BibitemShut {NoStop}%
\bibitem [{\citenamefont {Tucker-Smith}\ and\ \citenamefont
  {Weiner}(2001)}]{TuckerSmith:2001hy}%
  \BibitemOpen
  \bibfield  {author} {\bibinfo {author} {\bibfnamefont {D.}~\bibnamefont
  {Tucker-Smith}}\ and\ \bibinfo {author} {\bibfnamefont {N.}~\bibnamefont
  {Weiner}},\ }\href {\doibase 10.1103/PhysRevD.64.043502} {\bibfield
  {journal} {\bibinfo  {journal} {Phys. Rev.}\ }\textbf {\bibinfo {volume}
  {D64}},\ \bibinfo {pages} {043502} (\bibinfo {year} {2001})},\ \Eprint
  {http://arxiv.org/abs/hep-ph/0101138} {arXiv:hep-ph/0101138 [hep-ph]}
  \BibitemShut {NoStop}%
\bibitem [{\citenamefont {Finkbeiner}\ \emph {et~al.}(2009)\citenamefont
  {Finkbeiner}, \citenamefont {Lin},\ and\ \citenamefont
  {Weiner}}]{Finkbeiner:2009ug}%
  \BibitemOpen
  \bibfield  {author} {\bibinfo {author} {\bibfnamefont {D.~P.}\ \bibnamefont
  {Finkbeiner}}, \bibinfo {author} {\bibfnamefont {T.}~\bibnamefont {Lin}}, \
  and\ \bibinfo {author} {\bibfnamefont {N.}~\bibnamefont {Weiner}},\ }\href
  {\doibase 10.1103/PhysRevD.80.115008} {\bibfield  {journal} {\bibinfo
  {journal} {Phys. Rev.}\ }\textbf {\bibinfo {volume} {D80}},\ \bibinfo {pages}
  {115008} (\bibinfo {year} {2009})},\ \Eprint {http://arxiv.org/abs/0906.0002}
  {arXiv:0906.0002 [astro-ph.CO]} \BibitemShut {NoStop}%
\bibitem [{\citenamefont {Chang}\ \emph {et~al.}(2009)\citenamefont {Chang},
  \citenamefont {Kribs}, \citenamefont {Tucker-Smith},\ and\ \citenamefont
  {Weiner}}]{Chang:2008gd}%
  \BibitemOpen
  \bibfield  {author} {\bibinfo {author} {\bibfnamefont {S.}~\bibnamefont
  {Chang}}, \bibinfo {author} {\bibfnamefont {G.~D.}\ \bibnamefont {Kribs}},
  \bibinfo {author} {\bibfnamefont {D.}~\bibnamefont {Tucker-Smith}}, \ and\
  \bibinfo {author} {\bibfnamefont {N.}~\bibnamefont {Weiner}},\ }\href
  {\doibase 10.1103/PhysRevD.79.043513} {\bibfield  {journal} {\bibinfo
  {journal} {Phys. Rev.}\ }\textbf {\bibinfo {volume} {D79}},\ \bibinfo {pages}
  {043513} (\bibinfo {year} {2009})},\ \Eprint {http://arxiv.org/abs/0807.2250}
  {arXiv:0807.2250 [hep-ph]} \BibitemShut {NoStop}%
\bibitem [{\citenamefont {Menon}\ \emph {et~al.}(2010)\citenamefont {Menon},
  \citenamefont {Morris}, \citenamefont {Pierce},\ and\ \citenamefont
  {Weiner}}]{Menon:2009qj}%
  \BibitemOpen
  \bibfield  {author} {\bibinfo {author} {\bibfnamefont {A.}~\bibnamefont
  {Menon}}, \bibinfo {author} {\bibfnamefont {R.}~\bibnamefont {Morris}},
  \bibinfo {author} {\bibfnamefont {A.}~\bibnamefont {Pierce}}, \ and\ \bibinfo
  {author} {\bibfnamefont {N.}~\bibnamefont {Weiner}},\ }\href {\doibase
  10.1103/PhysRevD.82.015011} {\bibfield  {journal} {\bibinfo  {journal} {Phys.
  Rev.}\ }\textbf {\bibinfo {volume} {D82}},\ \bibinfo {pages} {015011}
  (\bibinfo {year} {2010})},\ \Eprint {http://arxiv.org/abs/0905.1847}
  {arXiv:0905.1847 [hep-ph]} \BibitemShut {NoStop}%
\bibitem [{\citenamefont {Nussinov}\ \emph {et~al.}(2009)\citenamefont
  {Nussinov}, \citenamefont {Wang},\ and\ \citenamefont
  {Yavin}}]{Nussinov:2009ft}%
  \BibitemOpen
  \bibfield  {author} {\bibinfo {author} {\bibfnamefont {S.}~\bibnamefont
  {Nussinov}}, \bibinfo {author} {\bibfnamefont {L.-T.}\ \bibnamefont {Wang}},
  \ and\ \bibinfo {author} {\bibfnamefont {I.}~\bibnamefont {Yavin}},\ }\href
  {\doibase 10.1088/1475-7516/2009/08/037} {\bibfield  {journal} {\bibinfo
  {journal} {JCAP}\ }\textbf {\bibinfo {volume} {0908}},\ \bibinfo {pages}
  {037} (\bibinfo {year} {2009})},\ \Eprint {http://arxiv.org/abs/0905.1333}
  {arXiv:0905.1333 [hep-ph]} \BibitemShut {NoStop}%
\bibitem [{\citenamefont {Catena}\ and\ \citenamefont
  {Hellstrom}(2018)}]{Catena:2018vzc}%
  \BibitemOpen
  \bibfield  {author} {\bibinfo {author} {\bibfnamefont {R.}~\bibnamefont
  {Catena}}\ and\ \bibinfo {author} {\bibfnamefont {F.}~\bibnamefont
  {Hellstrom}},\ }\href {\doibase 10.1088/1475-7516/2018/10/039} {\bibfield
  {journal} {\bibinfo  {journal} {JCAP}\ }\textbf {\bibinfo {volume} {1810}},\
  \bibinfo {pages} {039} (\bibinfo {year} {2018})},\ \Eprint
  {http://arxiv.org/abs/1808.08082} {arXiv:1808.08082 [astro-ph.CO]}
  \BibitemShut {NoStop}%
\bibitem [{\citenamefont {Blennow}\ \emph {et~al.}(2018)\citenamefont
  {Blennow}, \citenamefont {Clementz},\ and\ \citenamefont
  {Herrero-Garcia}}]{Blennow:2018xwu}%
  \BibitemOpen
  \bibfield  {author} {\bibinfo {author} {\bibfnamefont {M.}~\bibnamefont
  {Blennow}}, \bibinfo {author} {\bibfnamefont {S.}~\bibnamefont {Clementz}}, \
  and\ \bibinfo {author} {\bibfnamefont {J.}~\bibnamefont {Herrero-Garcia}},\
  }\href {\doibase 10.1140/epjc/s10052-018-5863-4} {\bibfield  {journal}
  {\bibinfo  {journal} {Eur. Phys. J.}\ }\textbf {\bibinfo {volume} {C78}},\
  \bibinfo {pages} {386} (\bibinfo {year} {2018})},\ \Eprint
  {http://arxiv.org/abs/1802.06880} {arXiv:1802.06880 [hep-ph]} \BibitemShut
  {NoStop}%
\bibitem [{\citenamefont {Abdo}\ \emph {et~al.}(2011)\citenamefont {Abdo} \emph
  {et~al.}}]{Abdo:2011xn}%
  \BibitemOpen
  \bibfield  {author} {\bibinfo {author} {\bibfnamefont {A.~A.}\ \bibnamefont
  {Abdo}} \emph {et~al.} (\bibinfo {collaboration} {Fermi-LAT}),\ }\href
  {\doibase 10.1088/0004-637X/734/2/116} {\bibfield  {journal} {\bibinfo
  {journal} {Astrophys. J.}\ }\textbf {\bibinfo {volume} {734}},\ \bibinfo
  {pages} {116} (\bibinfo {year} {2011})},\ \Eprint
  {http://arxiv.org/abs/1104.2093} {arXiv:1104.2093 [astro-ph.HE]} \BibitemShut
  {NoStop}%
\bibitem [{\citenamefont {Tang}\ \emph {et~al.}(2018)\citenamefont {Tang},
  \citenamefont {Ng}, \citenamefont {Linden}, \citenamefont {Zhou},
  \citenamefont {Beacom},\ and\ \citenamefont {Peter}}]{Tang:2018wqp}%
  \BibitemOpen
  \bibfield  {author} {\bibinfo {author} {\bibfnamefont {Q.-W.}\ \bibnamefont
  {Tang}}, \bibinfo {author} {\bibfnamefont {K.~C.~Y.}\ \bibnamefont {Ng}},
  \bibinfo {author} {\bibfnamefont {T.}~\bibnamefont {Linden}}, \bibinfo
  {author} {\bibfnamefont {B.}~\bibnamefont {Zhou}}, \bibinfo {author}
  {\bibfnamefont {J.~F.}\ \bibnamefont {Beacom}}, \ and\ \bibinfo {author}
  {\bibfnamefont {A.~H.~G.}\ \bibnamefont {Peter}},\ }\href {\doibase
  10.1103/PhysRevD.98.063019} {\bibfield  {journal} {\bibinfo  {journal} {Phys.
  Rev.}\ }\textbf {\bibinfo {volume} {D98}},\ \bibinfo {pages} {063019}
  (\bibinfo {year} {2018})},\ \Eprint {http://arxiv.org/abs/1804.06846}
  {arXiv:1804.06846 [astro-ph.HE]} \BibitemShut {NoStop}%
\bibitem [{\citenamefont {Seckel}\ \emph {et~al.}(1991)\citenamefont {Seckel},
  \citenamefont {Stanev},\ and\ \citenamefont {Gaisser}}]{Seckel:1991ffa}%
  \BibitemOpen
  \bibfield  {author} {\bibinfo {author} {\bibfnamefont {D.}~\bibnamefont
  {Seckel}}, \bibinfo {author} {\bibfnamefont {T.}~\bibnamefont {Stanev}}, \
  and\ \bibinfo {author} {\bibfnamefont {T.~K.}\ \bibnamefont {Gaisser}},\
  }\href {\doibase 10.1086/170753} {\bibfield  {journal} {\bibinfo  {journal}
  {Astrophys. J.}\ }\textbf {\bibinfo {volume} {382}},\ \bibinfo {pages} {652}
  (\bibinfo {year} {1991})}\BibitemShut {NoStop}%
\bibitem [{\citenamefont {Orlando}\ and\ \citenamefont
  {Strong}(2008)}]{Orlando:2008uk}%
  \BibitemOpen
  \bibfield  {author} {\bibinfo {author} {\bibfnamefont {E.}~\bibnamefont
  {Orlando}}\ and\ \bibinfo {author} {\bibfnamefont {A.~W.}\ \bibnamefont
  {Strong}},\ }\href {\doibase 10.1051/0004-6361:20078817} {\bibfield
  {journal} {\bibinfo  {journal} {Astron. Astrophys.}\ }\textbf {\bibinfo
  {volume} {480}},\ \bibinfo {pages} {847} (\bibinfo {year} {2008})},\ \Eprint
  {http://arxiv.org/abs/0801.2178} {arXiv:0801.2178 [astro-ph]} \BibitemShut
  {NoStop}%
\bibitem [{\citenamefont {Orlando}\ and\ \citenamefont
  {Strong}(2007)}]{Orlando:2006zs}%
  \BibitemOpen
  \bibfield  {author} {\bibinfo {author} {\bibfnamefont {E.}~\bibnamefont
  {Orlando}}\ and\ \bibinfo {author} {\bibfnamefont {A.}~\bibnamefont
  {Strong}},\ }\bibfield  {booktitle} {\emph {\bibinfo {booktitle} {{The
  Multi-Messenger Approach to High-Energy Gamma-Ray Sources: 3rd Workshop on
  the Nature of Unidentified High-Energy Sources, Barcelona, Spain, 4-7 Jul,
  2006}}},\ }\href {\doibase 10.1007/s10509-007-9457-0} {\bibfield  {journal}
  {\bibinfo  {journal} {Astrophys. Space Sci.}\ }\textbf {\bibinfo {volume}
  {309}},\ \bibinfo {pages} {359} (\bibinfo {year} {2007})},\ \Eprint
  {http://arxiv.org/abs/astro-ph/0607563} {arXiv:astro-ph/0607563 [astro-ph]}
  \BibitemShut {NoStop}%
\bibitem [{\citenamefont {Mazziotta}\ \emph {et~al.}(2020)\citenamefont
  {Mazziotta}, \citenamefont {De~La Torre~Luque}, \citenamefont {Di~Venere},
  \citenamefont {Fass{\`o}}, \citenamefont {Ferrari}, \citenamefont {Loparco},
  \citenamefont {Sala},\ and\ \citenamefont {Serini}}]{mazziotta2020cosmic}%
  \BibitemOpen
  \bibfield  {author} {\bibinfo {author} {\bibfnamefont {M.}~\bibnamefont
  {Mazziotta}}, \bibinfo {author} {\bibfnamefont {P.}~\bibnamefont {De~La
  Torre~Luque}}, \bibinfo {author} {\bibfnamefont {L.}~\bibnamefont
  {Di~Venere}}, \bibinfo {author} {\bibfnamefont {A.}~\bibnamefont
  {Fass{\`o}}}, \bibinfo {author} {\bibfnamefont {A.}~\bibnamefont {Ferrari}},
  \bibinfo {author} {\bibfnamefont {F.}~\bibnamefont {Loparco}}, \bibinfo
  {author} {\bibfnamefont {P.}~\bibnamefont {Sala}}, \ and\ \bibinfo {author}
  {\bibfnamefont {D.}~\bibnamefont {Serini}},\ }\href {\doibase
  10.1103/PhysRevD.101.083011} {\bibfield  {journal} {\bibinfo  {journal}
  {Phys. Rev. D}\ }\textbf {\bibinfo {volume} {101}},\ \bibinfo {pages}
  {083011} (\bibinfo {year} {2020})},\ \Eprint
  {http://arxiv.org/abs/2001.09933} {arXiv:2001.09933 [astro-ph.HE]}
  \BibitemShut {NoStop}%
\bibitem [{\citenamefont {Cuoco}\ \emph {et~al.}(2020)\citenamefont {Cuoco},
  \citenamefont {De~La Torre~Luque}, \citenamefont {Gargano}, \citenamefont
  {Gustafsson}, \citenamefont {Loparco}, \citenamefont {Mazziotta},\ and\
  \citenamefont {Serini}}]{Mazziotta:2019CREs}%
  \BibitemOpen
  \bibfield  {author} {\bibinfo {author} {\bibfnamefont {A.}~\bibnamefont
  {Cuoco}}, \bibinfo {author} {\bibfnamefont {P.}~\bibnamefont {De~La
  Torre~Luque}}, \bibinfo {author} {\bibfnamefont {F.}~\bibnamefont {Gargano}},
  \bibinfo {author} {\bibfnamefont {M.}~\bibnamefont {Gustafsson}}, \bibinfo
  {author} {\bibfnamefont {F.}~\bibnamefont {Loparco}}, \bibinfo {author}
  {\bibfnamefont {M.~N.}\ \bibnamefont {Mazziotta}}, \ and\ \bibinfo {author}
  {\bibfnamefont {D.}~\bibnamefont {Serini}},\ }\href {\doibase
  10.1103/PhysRevD.101.022002} {\bibfield  {journal} {\bibinfo  {journal}
  {Phys. Rev.}\ }\textbf {\bibinfo {volume} {D101}},\ \bibinfo {pages} {022002}
  (\bibinfo {year} {2020})},\ \Eprint {http://arxiv.org/abs/1912.09373}
  {arXiv:1912.09373 [astro-ph.HE]} \BibitemShut {NoStop}%
\bibitem [{\citenamefont {Ibarra}\ \emph {et~al.}(2012)\citenamefont {Ibarra},
  \citenamefont {Lopez~Gehler},\ and\ \citenamefont {Pato}}]{Ibarra:2012dw}%
  \BibitemOpen
  \bibfield  {author} {\bibinfo {author} {\bibfnamefont {A.}~\bibnamefont
  {Ibarra}}, \bibinfo {author} {\bibfnamefont {S.}~\bibnamefont
  {Lopez~Gehler}}, \ and\ \bibinfo {author} {\bibfnamefont {M.}~\bibnamefont
  {Pato}},\ }\href {\doibase 10.1088/1475-7516/2012/07/043} {\bibfield
  {journal} {\bibinfo  {journal} {JCAP}\ }\textbf {\bibinfo {volume} {1207}},\
  \bibinfo {pages} {043} (\bibinfo {year} {2012})},\ \Eprint
  {http://arxiv.org/abs/1205.0007} {arXiv:1205.0007 [hep-ph]} \BibitemShut
  {NoStop}%
\bibitem [{\citenamefont {Sivertsson}\ and\ \citenamefont
  {Edsjo}(2010)}]{Sivertsson:2009nx}%
  \BibitemOpen
  \bibfield  {author} {\bibinfo {author} {\bibfnamefont {S.}~\bibnamefont
  {Sivertsson}}\ and\ \bibinfo {author} {\bibfnamefont {J.}~\bibnamefont
  {Edsjo}},\ }\href {\doibase 10.1103/PhysRevD.81.063502} {\bibfield  {journal}
  {\bibinfo  {journal} {Phys. Rev.}\ }\textbf {\bibinfo {volume} {D81}},\
  \bibinfo {pages} {063502} (\bibinfo {year} {2010})},\ \Eprint
  {http://arxiv.org/abs/0910.0017} {arXiv:0910.0017 [astro-ph.HE]} \BibitemShut
  {NoStop}%
\bibitem [{\citenamefont {Albert}\ \emph {et~al.}(2018)\citenamefont {Albert}
  \emph {et~al.}}]{Albert:2018jwh}%
  \BibitemOpen
  \bibfield  {author} {\bibinfo {author} {\bibfnamefont {A.}~\bibnamefont
  {Albert}} \emph {et~al.} (\bibinfo {collaboration} {HAWC}),\ }\href {\doibase
  10.1103/PhysRevD.98.123012} {\bibfield  {journal} {\bibinfo  {journal} {Phys.
  Rev. D}\ } (\bibinfo {year} {2018}),\ 10.1103/PhysRevD.98.123012},\ \bibinfo
  {note} {[Phys. Rev.D98,123012(2018)]},\ \Eprint
  {http://arxiv.org/abs/1808.05624} {arXiv:1808.05624 [hep-ph]} \BibitemShut
  {NoStop}%
\bibitem [{\citenamefont {Griest}\ and\ \citenamefont
  {Seckel}(1987{\natexlab{a}})}]{griest1987cosmic}%
  \BibitemOpen
  \bibfield  {author} {\bibinfo {author} {\bibfnamefont {K.}~\bibnamefont
  {Griest}}\ and\ \bibinfo {author} {\bibfnamefont {D.}~\bibnamefont
  {Seckel}},\ }\href {\doibase https://doi.org/10.1016/0550-3213(87)90293-8}
  {\bibfield  {journal} {\bibinfo  {journal} {Nuclear Physics B}\ }\textbf
  {\bibinfo {volume} {283}},\ \bibinfo {pages} {681} (\bibinfo {year}
  {1987}{\natexlab{a}})}\BibitemShut {NoStop}%
\bibitem [{\citenamefont {Jungman}\ \emph {et~al.}(1996)\citenamefont
  {Jungman}, \citenamefont {Kamionkowski},\ and\ \citenamefont
  {Griest}}]{Jungman:1995df}%
  \BibitemOpen
  \bibfield  {author} {\bibinfo {author} {\bibfnamefont {G.}~\bibnamefont
  {Jungman}}, \bibinfo {author} {\bibfnamefont {M.}~\bibnamefont
  {Kamionkowski}}, \ and\ \bibinfo {author} {\bibfnamefont {K.}~\bibnamefont
  {Griest}},\ }\href {\doibase 10.1016/0370-1573(95)00058-5} {\bibfield
  {journal} {\bibinfo  {journal} {Phys. Rept.}\ }\textbf {\bibinfo {volume}
  {267}},\ \bibinfo {pages} {195} (\bibinfo {year} {1996})},\ \Eprint
  {http://arxiv.org/abs/hep-ph/9506380} {arXiv:hep-ph/9506380 [hep-ph]}
  \BibitemShut {NoStop}%
\bibitem [{\citenamefont {Gondolo}\ \emph {et~al.}(2004)\citenamefont
  {Gondolo}, \citenamefont {Edsjo}, \citenamefont {Ullio}, \citenamefont
  {Bergstrom}, \citenamefont {Schelke},\ and\ \citenamefont
  {Baltz}}]{Gondolo:2004sc}%
  \BibitemOpen
  \bibfield  {author} {\bibinfo {author} {\bibfnamefont {P.}~\bibnamefont
  {Gondolo}}, \bibinfo {author} {\bibfnamefont {J.}~\bibnamefont {Edsjo}},
  \bibinfo {author} {\bibfnamefont {P.}~\bibnamefont {Ullio}}, \bibinfo
  {author} {\bibfnamefont {L.}~\bibnamefont {Bergstrom}}, \bibinfo {author}
  {\bibfnamefont {M.}~\bibnamefont {Schelke}}, \ and\ \bibinfo {author}
  {\bibfnamefont {E.~A.}\ \bibnamefont {Baltz}},\ }\href {\doibase
  10.1088/1475-7516/2004/07/008} {\bibfield  {journal} {\bibinfo  {journal}
  {JCAP}\ }\textbf {\bibinfo {volume} {0407}},\ \bibinfo {pages} {008}
  (\bibinfo {year} {2004})},\ \Eprint {http://arxiv.org/abs/astro-ph/0406204}
  {arXiv:astro-ph/0406204 [astro-ph]} \BibitemShut {NoStop}%
\bibitem [{\citenamefont {Bringmann}\ \emph {et~al.}(2018)\citenamefont
  {Bringmann}, \citenamefont {Edsjö}, \citenamefont {Gondolo}, \citenamefont
  {Ullio},\ and\ \citenamefont {Bergström}}]{Bringmann:2018lay}%
  \BibitemOpen
  \bibfield  {author} {\bibinfo {author} {\bibfnamefont {T.}~\bibnamefont
  {Bringmann}}, \bibinfo {author} {\bibfnamefont {J.}~\bibnamefont {Edsjö}},
  \bibinfo {author} {\bibfnamefont {P.}~\bibnamefont {Gondolo}}, \bibinfo
  {author} {\bibfnamefont {P.}~\bibnamefont {Ullio}}, \ and\ \bibinfo {author}
  {\bibfnamefont {L.}~\bibnamefont {Bergström}},\ }\href {\doibase
  10.1088/1475-7516/2018/07/033} {\bibfield  {journal} {\bibinfo  {journal}
  {JCAP}\ }\textbf {\bibinfo {volume} {1807}},\ \bibinfo {pages} {033}
  (\bibinfo {year} {2018})},\ \Eprint {http://arxiv.org/abs/1802.03399}
  {arXiv:1802.03399 [hep-ph]} \BibitemShut {NoStop}%
\bibitem [{\citenamefont {Edsjo}\ \emph {et~al.}()\citenamefont {Edsjo},
  \citenamefont {Bringmann}, \citenamefont {Gondolo}, , \citenamefont {Ullio},
  \citenamefont {Bergstrom}, \citenamefont {Schelke}, \citenamefont {Baltz},\
  and\ \citenamefont {Duda}}]{darksusyweb}%
  \BibitemOpen
  \bibfield  {author} {\bibinfo {author} {\bibfnamefont {J.}~\bibnamefont
  {Edsjo}}, \bibinfo {author} {\bibfnamefont {T.}~\bibnamefont {Bringmann}},
  \bibinfo {author} {\bibfnamefont {P.}~\bibnamefont {Gondolo}}, , \bibinfo
  {author} {\bibfnamefont {P.}~\bibnamefont {Ullio}}, \bibinfo {author}
  {\bibfnamefont {L.}~\bibnamefont {Bergstrom}}, \bibinfo {author}
  {\bibfnamefont {M.}~\bibnamefont {Schelke}}, \bibinfo {author} {\bibfnamefont
  {E.~A.}\ \bibnamefont {Baltz}}, \ and\ \bibinfo {author} {\bibfnamefont
  {G.}~\bibnamefont {Duda}},\ }\href@noop {} {}\bibinfo {howpublished}
  {\url{http://www.darksusy.org/}}\BibitemShut {NoStop}%
\bibitem [{\citenamefont {Atwood}\ \emph {et~al.}(2013)\citenamefont {Atwood}
  \emph {et~al.}}]{Atwood:2013rka}%
  \BibitemOpen
  \bibfield  {author} {\bibinfo {author} {\bibfnamefont {W.}~\bibnamefont
  {Atwood}} \emph {et~al.} (\bibinfo {collaboration} {Fermi-LAT}),\ }\href@noop
  {} {\  (\bibinfo {year} {2013})},\ \Eprint {http://arxiv.org/abs/1303.3514}
  {arXiv:1303.3514 [astro-ph.IM]} \BibitemShut {NoStop}%
\bibitem [{\citenamefont {Bruel}\ \emph {et~al.}(2018)\citenamefont {Bruel},
  \citenamefont {Burnett}, \citenamefont {Digel}, \citenamefont {Johannesson},
  \citenamefont {Omodei},\ and\ \citenamefont {Wood}}]{Bruel:2018lac}%
  \BibitemOpen
  \bibfield  {author} {\bibinfo {author} {\bibfnamefont {P.}~\bibnamefont
  {Bruel}}, \bibinfo {author} {\bibfnamefont {T.~H.}\ \bibnamefont {Burnett}},
  \bibinfo {author} {\bibfnamefont {S.~W.}\ \bibnamefont {Digel}}, \bibinfo
  {author} {\bibfnamefont {G.}~\bibnamefont {Johannesson}}, \bibinfo {author}
  {\bibfnamefont {N.}~\bibnamefont {Omodei}}, \ and\ \bibinfo {author}
  {\bibfnamefont {M.}~\bibnamefont {Wood}} (\bibinfo {collaboration}
  {Fermi-LAT}),\ }in\ \href@noop {} {\emph {\bibinfo {booktitle} {{8th
  International Fermi Symposium: Celebrating 10 Year of Fermi Baltimore,
  Maryland, USA, October 14-19, 2018}}}}\ (\bibinfo {year} {2018})\ \Eprint
  {http://arxiv.org/abs/1810.11394} {arXiv:1810.11394 [astro-ph.IM]}
  \BibitemShut {NoStop}%
\bibitem [{cic()}]{cicerone}%
  \BibitemOpen
  \href@noop {} {}\bibinfo {howpublished} {\url{
  https://fermi.gsfc.nasa.gov/ssc/data/analysis/documentation/Cicerone/Ciceron%
e_Data/LAT_DP.html}}\BibitemShut {NoStop}%
\bibitem [{\citenamefont {Acero}\ \emph {et~al.}(2015)\citenamefont {Acero}
  \emph {et~al.}}]{Acero:2015hja}%
  \BibitemOpen
  \bibfield  {author} {\bibinfo {author} {\bibfnamefont {F.}~\bibnamefont
  {Acero}} \emph {et~al.} (\bibinfo {collaboration} {Fermi-LAT}),\ }\href
  {\doibase 10.1088/0067-0049/218/2/23} {\bibfield  {journal} {\bibinfo
  {journal} {Astrophys. J. Suppl.}\ }\textbf {\bibinfo {volume} {218}},\
  \bibinfo {pages} {23} (\bibinfo {year} {2015})},\ \Eprint
  {http://arxiv.org/abs/1501.02003} {arXiv:1501.02003 [astro-ph.HE]}
  \BibitemShut {NoStop}%
\bibitem [{\citenamefont {Mazziotta}\ \emph {et~al.}(2018)\citenamefont
  {Mazziotta}, \citenamefont {Costanza}, \citenamefont {Cuoco}, \citenamefont
  {Gargano}, \citenamefont {Loparco},\ and\ \citenamefont
  {Zimmer}}]{Mazziotta:2017ruy}%
  \BibitemOpen
  \bibfield  {author} {\bibinfo {author} {\bibfnamefont {M.~N.}\ \bibnamefont
  {Mazziotta}}, \bibinfo {author} {\bibfnamefont {F.}~\bibnamefont {Costanza}},
  \bibinfo {author} {\bibfnamefont {A.}~\bibnamefont {Cuoco}}, \bibinfo
  {author} {\bibfnamefont {F.}~\bibnamefont {Gargano}}, \bibinfo {author}
  {\bibfnamefont {F.}~\bibnamefont {Loparco}}, \ and\ \bibinfo {author}
  {\bibfnamefont {S.}~\bibnamefont {Zimmer}},\ }\href {\doibase
  10.1103/PhysRevD.98.022006} {\bibfield  {journal} {\bibinfo  {journal} {Phys.
  Rev.}\ }\textbf {\bibinfo {volume} {D98}},\ \bibinfo {pages} {022006}
  (\bibinfo {year} {2018})},\ \Eprint {http://arxiv.org/abs/1712.07005}
  {arXiv:1712.07005 [astro-ph.HE]} \BibitemShut {NoStop}%
\bibitem [{\citenamefont {Mazziotta}(2009)}]{Mazziotta:2009rd}%
  \BibitemOpen
  \bibfield  {author} {\bibinfo {author} {\bibfnamefont {M.~N.}\ \bibnamefont
  {Mazziotta}} (\bibinfo {collaboration} {Fermi-LAT}),\ }\href@noop {} {\
  (\bibinfo {year} {2009})},\ \Eprint {http://arxiv.org/abs/0912.1236}
  {arXiv:0912.1236 [astro-ph.IM]} \BibitemShut {NoStop}%
\bibitem [{\citenamefont {Loparco}\ and\ \citenamefont
  {Mazziotta}(2009)}]{Loparco:2009by}%
  \BibitemOpen
  \bibfield  {author} {\bibinfo {author} {\bibfnamefont {F.}~\bibnamefont
  {Loparco}}\ and\ \bibinfo {author} {\bibfnamefont {M.~N.}\ \bibnamefont
  {Mazziotta}} (\bibinfo {collaboration} {Fermi-LAT}),\ }\href@noop {} {\
  (\bibinfo {year} {2009})},\ \Eprint {http://arxiv.org/abs/0912.3695}
  {arXiv:0912.3695 [astro-ph.IM]} \BibitemShut {NoStop}%
\bibitem [{\citenamefont {Ackermann}\ \emph {et~al.}(2012)\citenamefont
  {Ackermann} \emph {et~al.}}]{Ackermann:2012kna}%
  \BibitemOpen
  \bibfield  {author} {\bibinfo {author} {\bibfnamefont {M.}~\bibnamefont
  {Ackermann}} \emph {et~al.} (\bibinfo {collaboration} {Fermi-LAT}),\ }\href
  {\doibase 10.1088/0067-0049/203/1/4} {\bibfield  {journal} {\bibinfo
  {journal} {Astrophys. J. Suppl.}\ }\textbf {\bibinfo {volume} {203}},\
  \bibinfo {pages} {4} (\bibinfo {year} {2012})},\ \Eprint
  {http://arxiv.org/abs/1206.1896} {arXiv:1206.1896 [astro-ph.IM]} \BibitemShut
  {NoStop}%
\bibitem [{\citenamefont {Brun}\ and\ \citenamefont
  {Rademakers}(1997)}]{Brun:1997pa}%
  \BibitemOpen
  \bibfield  {author} {\bibinfo {author} {\bibfnamefont {R.}~\bibnamefont
  {Brun}}\ and\ \bibinfo {author} {\bibfnamefont {F.}~\bibnamefont
  {Rademakers}},\ }\bibfield  {booktitle} {\emph {\bibinfo {booktitle} {{New
  computing techniques in physics research V. Proceedings, 5th International
  Workshop, AIHENP '96, Lausanne, Switzerland, September 2-6, 1996}}},\ }\href
  {\doibase 10.1016/S0168-9002(97)00048-X} {\bibfield  {journal} {\bibinfo
  {journal} {Nucl. Instrum. Meth.}\ }\textbf {\bibinfo {volume} {A389}},\
  \bibinfo {pages} {81} (\bibinfo {year} {1997})}\BibitemShut {NoStop}%
\bibitem [{roo()}]{rootweb}%
  \BibitemOpen
  \href@noop {} {}\bibinfo {howpublished} {\url{https://root.cern.ch/} release
  5.34}\BibitemShut {NoStop}%
\bibitem [{P8p()}]{P8performance}%
  \BibitemOpen
  \href@noop {} {}\bibinfo {howpublished}
  {\url{http://www.slac.stanford.edu/exp/glast/groups/canda/lat_Performance.ht%
m}}\BibitemShut {NoStop}%
\bibitem [{\citenamefont {Aprile}\ \emph {et~al.}(2018)\citenamefont {Aprile}
  \emph {et~al.}}]{Aprile:2018dbl}%
  \BibitemOpen
  \bibfield  {author} {\bibinfo {author} {\bibfnamefont {E.}~\bibnamefont
  {Aprile}} \emph {et~al.} (\bibinfo {collaboration} {XENON}),\ }\href
  {\doibase 10.1103/PhysRevLett.121.111302} {\bibfield  {journal} {\bibinfo
  {journal} {Phys. Rev. Lett.}\ }\textbf {\bibinfo {volume} {121}},\ \bibinfo
  {pages} {111302} (\bibinfo {year} {2018})},\ \Eprint
  {http://arxiv.org/abs/1805.12562} {arXiv:1805.12562 [astro-ph.CO]}
  \BibitemShut {NoStop}%
\bibitem [{\citenamefont {Amole}\ \emph {et~al.}(2019)\citenamefont {Amole}
  \emph {et~al.}}]{Amole:2019fdf}%
  \BibitemOpen
  \bibfield  {author} {\bibinfo {author} {\bibfnamefont {C.}~\bibnamefont
  {Amole}} \emph {et~al.} (\bibinfo {collaboration} {PICO}),\ }\href {\doibase
  10.1103/PhysRevD.100.022001} {\bibfield  {journal} {\bibinfo  {journal}
  {Phys. Rev.}\ }\textbf {\bibinfo {volume} {D100}},\ \bibinfo {pages} {022001}
  (\bibinfo {year} {2019})},\ \Eprint {http://arxiv.org/abs/1902.04031}
  {arXiv:1902.04031 [astro-ph.CO]} \BibitemShut {NoStop}%
\bibitem [{\citenamefont {Griest}\ and\ \citenamefont
  {Seckel}(1987{\natexlab{b}})}]{Griest:1986yu}%
  \BibitemOpen
  \bibfield  {author} {\bibinfo {author} {\bibfnamefont {K.}~\bibnamefont
  {Griest}}\ and\ \bibinfo {author} {\bibfnamefont {D.}~\bibnamefont
  {Seckel}},\ }\href {\doibase 10.1016/0550-3213(87)90293-8,
  10.1016/0550-3213(88)90409-9} {\bibfield  {journal} {\bibinfo  {journal}
  {Nucl. Phys.}\ }\textbf {\bibinfo {volume} {B283}},\ \bibinfo {pages} {681}
  (\bibinfo {year} {1987}{\natexlab{b}})},\ \bibinfo {note} {[Erratum: Nucl.
  Phys.B296,1034(1988)]}\BibitemShut {NoStop}%
\bibitem [{\citenamefont {Steigman}\ \emph {et~al.}(2012)\citenamefont
  {Steigman}, \citenamefont {Dasgupta},\ and\ \citenamefont
  {Beacom}}]{Steigman:2012nb}%
  \BibitemOpen
  \bibfield  {author} {\bibinfo {author} {\bibfnamefont {G.}~\bibnamefont
  {Steigman}}, \bibinfo {author} {\bibfnamefont {B.}~\bibnamefont {Dasgupta}},
  \ and\ \bibinfo {author} {\bibfnamefont {J.~F.}\ \bibnamefont {Beacom}},\
  }\href {\doibase 10.1103/PhysRevD.86.023506} {\bibfield  {journal} {\bibinfo
  {journal} {Phys. Rev. D}\ }\textbf {\bibinfo {volume} {86}},\ \bibinfo
  {pages} {023506} (\bibinfo {year} {2012})},\ \Eprint
  {http://arxiv.org/abs/1204.3622} {arXiv:1204.3622 [hep-ph]} \BibitemShut
  {NoStop}%
\bibitem [{\citenamefont {Ackermann}\ \emph {et~al.}(2015)\citenamefont
  {Ackermann} \emph {et~al.}}]{Ackermann:2015lka}%
  \BibitemOpen
  \bibfield  {author} {\bibinfo {author} {\bibfnamefont {M.}~\bibnamefont
  {Ackermann}} \emph {et~al.} (\bibinfo {collaboration} {Fermi-LAT}),\ }\href
  {\doibase 10.1103/PhysRevD.91.122002} {\bibfield  {journal} {\bibinfo
  {journal} {Phys. Rev.}\ }\textbf {\bibinfo {volume} {D91}},\ \bibinfo {pages}
  {122002} (\bibinfo {year} {2015})},\ \Eprint
  {http://arxiv.org/abs/1506.00013} {arXiv:1506.00013 [astro-ph.HE]}
  \BibitemShut {NoStop}%
\bibitem [{\citenamefont {Depta}\ \emph {et~al.}(2019)\citenamefont {Depta},
  \citenamefont {Hufnagel}, \citenamefont {Schmidt-Hoberg},\ and\ \citenamefont
  {Wild}}]{Depta:2019lbe}%
  \BibitemOpen
  \bibfield  {author} {\bibinfo {author} {\bibfnamefont {P.~F.}\ \bibnamefont
  {Depta}}, \bibinfo {author} {\bibfnamefont {M.}~\bibnamefont {Hufnagel}},
  \bibinfo {author} {\bibfnamefont {K.}~\bibnamefont {Schmidt-Hoberg}}, \ and\
  \bibinfo {author} {\bibfnamefont {S.}~\bibnamefont {Wild}},\ }\href {\doibase
  10.1088/1475-7516/2019/04/029} {\bibfield  {journal} {\bibinfo  {journal}
  {JCAP}\ }\textbf {\bibinfo {volume} {04}},\ \bibinfo {pages} {029} (\bibinfo
  {year} {2019})},\ \Eprint {http://arxiv.org/abs/1901.06944} {arXiv:1901.06944
  [hep-ph]} \BibitemShut {NoStop}%
\bibitem [{\citenamefont {Jedamzik}\ and\ \citenamefont
  {Pospelov}(2009)}]{Jedamzik:2009uy}%
  \BibitemOpen
  \bibfield  {author} {\bibinfo {author} {\bibfnamefont {K.}~\bibnamefont
  {Jedamzik}}\ and\ \bibinfo {author} {\bibfnamefont {M.}~\bibnamefont
  {Pospelov}},\ }\href {\doibase 10.1088/1367-2630/11/10/105028} {\bibfield
  {journal} {\bibinfo  {journal} {New J. Phys.}\ }\textbf {\bibinfo {volume}
  {11}},\ \bibinfo {pages} {105028} (\bibinfo {year} {2009})},\ \Eprint
  {http://arxiv.org/abs/0906.2087} {arXiv:0906.2087 [hep-ph]} \BibitemShut
  {NoStop}%
\bibitem [{\citenamefont {Slatyer}(2016)}]{Slatyer:2015jla}%
  \BibitemOpen
  \bibfield  {author} {\bibinfo {author} {\bibfnamefont {T.~R.}\ \bibnamefont
  {Slatyer}},\ }\href {\doibase 10.1103/PhysRevD.93.023527} {\bibfield
  {journal} {\bibinfo  {journal} {Phys. Rev. D}\ }\textbf {\bibinfo {volume}
  {93}},\ \bibinfo {pages} {023527} (\bibinfo {year} {2016})},\ \Eprint
  {http://arxiv.org/abs/1506.03811} {arXiv:1506.03811 [hep-ph]} \BibitemShut
  {NoStop}%
\bibitem [{\citenamefont {Ibarra}\ \emph {et~al.}(2013)\citenamefont {Ibarra},
  \citenamefont {Lee}, \citenamefont {López~Gehler}, \citenamefont {Park},\
  and\ \citenamefont {Pato}}]{Ibarra:2013eda}%
  \BibitemOpen
  \bibfield  {author} {\bibinfo {author} {\bibfnamefont {A.}~\bibnamefont
  {Ibarra}}, \bibinfo {author} {\bibfnamefont {H.~M.}\ \bibnamefont {Lee}},
  \bibinfo {author} {\bibfnamefont {S.}~\bibnamefont {López~Gehler}}, \bibinfo
  {author} {\bibfnamefont {W.-I.}\ \bibnamefont {Park}}, \ and\ \bibinfo
  {author} {\bibfnamefont {M.}~\bibnamefont {Pato}},\ }\href {\doibase
  10.1088/1475-7516/2016/03/E01, 10.1088/1475-7516/2013/05/016} {\bibfield
  {journal} {\bibinfo  {journal} {JCAP}\ }\textbf {\bibinfo {volume} {1305}},\
  \bibinfo {pages} {016} (\bibinfo {year} {2013})},\ \bibinfo {note} {[Erratum:
  JCAP1603,no.03,E01(2016)]},\ \Eprint {http://arxiv.org/abs/1303.6632}
  {arXiv:1303.6632 [hep-ph]} \BibitemShut {NoStop}%
\bibitem [{\citenamefont {Li}\ \emph {et~al.}(2018)\citenamefont {Li},
  \citenamefont {Liang}, \citenamefont {Xia}, \citenamefont {Zu}, \citenamefont
  {Duan}, \citenamefont {Shen}, \citenamefont {Feng}, \citenamefont {Yuan},\
  and\ \citenamefont {Fan}}]{Li:2018rqo}%
  \BibitemOpen
  \bibfield  {author} {\bibinfo {author} {\bibfnamefont {S.}~\bibnamefont
  {Li}}, \bibinfo {author} {\bibfnamefont {Y.-F.}\ \bibnamefont {Liang}},
  \bibinfo {author} {\bibfnamefont {Z.-Q.}\ \bibnamefont {Xia}}, \bibinfo
  {author} {\bibfnamefont {L.}~\bibnamefont {Zu}}, \bibinfo {author}
  {\bibfnamefont {K.-K.}\ \bibnamefont {Duan}}, \bibinfo {author}
  {\bibfnamefont {Z.-Q.}\ \bibnamefont {Shen}}, \bibinfo {author}
  {\bibfnamefont {L.}~\bibnamefont {Feng}}, \bibinfo {author} {\bibfnamefont
  {Q.}~\bibnamefont {Yuan}}, \ and\ \bibinfo {author} {\bibfnamefont {Y.-Z.}\
  \bibnamefont {Fan}},\ }\href {\doibase 10.1103/PhysRevD.97.083007} {\bibfield
   {journal} {\bibinfo  {journal} {Phys. Rev.}\ }\textbf {\bibinfo {volume}
  {D97}},\ \bibinfo {pages} {083007} (\bibinfo {year} {2018})}\BibitemShut
  {NoStop}%
\end{thebibliography}%

\end{document}